\documentclass{aa}
\usepackage{graphicx}
\usepackage{txfonts}
\usepackage{longtable}
\usepackage{natbib}  % citet, citep, citealp
\bibpunct{(}{)}{;}{a}{}{,}

\begin{document}

   \title{Survey of Planetary Nebulae at 30\,GHz with OCRA-p }

   \author{B.\,M.~Pazderska
          \inst{1}
          \and
          M.\,P.~Gawro\'nski
\inst{1}
\and
R.~Feiler
\inst{1}
\and
M.~Birkinshaw
\inst{3}
\and
I.\,W.\,A.~Browne
\inst{2}
\and
R.~Davis
\inst{2}
\and
A.\,J.~Kus
\inst{1}
\and
K.~Lancaster
\inst{3}
\and
S.\,R.~Lowe
\inst{2}
\and
E.~Pazderski
\inst{1}
\and
M.~Peel
\inst{2}
\and
P.\,N.~Wilkinson
\inst{2}
}

\offprints{B.~Pazderska, e-mail: bogna@epsrv.astro.uni.torun.pl}
   \institute{Toru\'n Centre for Astronomy, Nicolaus Copernicus University, 87-100 Toru\'n/Piwnice, Poland\\
         \and
      Jodrell Bank Centre for Astrophysics, University of Manchaster, Manchester, M13 9PL, UK\\
	\and
      University of Bristol, Tyndall Avenue, Bristol BS8 ITL, UK\\
       }

   \date{Received November 18, 2008;  accepted February 18, 2009}
 
  \abstract
   {}
   {We report the results of a survey of 442 planetary nebulae at 30\,GHz. The purpose of the survey is to develop a list of planetary nebulae as calibration sources which could be used for high frequency calibration in future. For 41 PNe with sufficient data, we test the emission mechanisms in order to evaluate whether or not spinning dust plays an important role in their spectra at 30\,GHz.}
   {The 30-GHz data were obtained with a twin-beam differencing radiometer, OCRA-p, which is in operation on the Toru\'n 32-m telescope. Sources were scanned both in right ascension and declination. We estimated flux densities at 30\,GHz using a free-free emission model and compared it with our data.}
   {The primary result is a catalogue containing the flux densities of 93 planetary nebulae at 30\,GHz. Sources with sufficient data were compared with a spectral model of free-free emission. The model shows that free-free emission can generally explain the observed flux densities at 30\,GHz thus no other emission mechanism is needed to account for the high frequency spectra.}
   {}


   \maketitle
%
%________________________________________________________________

\section{Introduction}
 The planetary nebula (PN) phase in the evolution of low mass stars lasts only about $10^4$ years. It begins once the central star reaches an effective temperature of 20\,000\,K and ionises the shell of material developed during Asymptotic Giant Branch (AGB) evolution. The end of this phase is defined by termination of nuclear burning in a thin outer shell of the star, and then rapid dispersal of the nebula. At radio wavelengths, planetary nebulae emit continuum radiation in a free-free process and are among the brightest Galactic radio sources. Their radio flux densities do not suffer the high levels of extinction present in the optical regime. Since most planetary nebulae occupy Galactic plane where extinction is high, radio detections and radio flux density measurements are important.

Planetary nebulae are mostly compact sources because they are distant or intrinsically small. Their relatively strong and stable radio emission makes them good candidates for calibration sources.

Many radio continuum observations of planetary nebulae have been made at 1.4, 5 and 14.7\,GHz \citep[see][]{CondonCAT,AckerCAT,Aaquist,Milne}. There is still only limited data at frequencies above 30\,GHz and most of those observations were obtained with interferometers such as the VLA; relatively little comes from single dishes. To extend the spectral range and to make total flux density measurements, we have used the new One Centimetre Radio Array prototype (OCRA-p) receiver to observe planetary nebulae at 30\,GHz. This receiver is mounted on Toru\'n's 32-m radio telescope and is described in section \ref{ocra}. OCRA-p receiver is outlined in detail by \citet{Lowe1}. The survey for planetary nebulae was one of the first successful observations made using this system (together with measurements of flat-spectrum sources by \citealp{Lowe}, and observations of the Sunayaev Zel'dovich effect by \citealp{Lancaster}). The purpose was to make a list of high frequency calibrators, which can be used to support sky surveys and to test the emission mechanisms in order to evaluate whether or not spinning dust plays an important role in PN spectra.

Our new survey of planetary nebulae brought detections of 93 sources at 30\,GHz out of 442, of which selection criteria are described in section~\ref{obs}. The observing techniques and data reduction process are described in section~\ref{obs}. The results are described in section~\ref{res}. The comparison of flux densities with the free-free emission model proposed by \citet{Siodmiak} is described in section~\ref{model}. Section~\ref{con} contains the final conclusions. The measured 30-GHz flux densities of all detected PNe are given in Table~\ref{table1}.

\section{OCRA-p}
\label{ocra}
OCRA-p is a 2-element prototype for OCRA \citep{Browne} - a 100-element array receiver. The OCRA-p receiver was constructed at the University of Manchester and was funded by a EU Faraday FP6 grant together with the Royal Society Paul Instrument fund. It has been mounted on the Toru\'n 32-m radio telescope owned by the Centre for Astronomy, Nicolaus Copernicus University, Toru\'n in Poland. 

%__________________________________________________ One column table
   \begin{table}
\label{table0}
      \caption[]{OCRA-p nominal system specification \citep[from][]{Lowe}}
         \label{KapSou}
     $$ 
         \begin{array}{p{0.5\linewidth}l}
            \hline
            \noalign{\smallskip}
            ~~~Frequency range    &  27-33\,\mathrm{GHz}~~ \\
           ~~~System temperature & 40\,\mathrm{K}~~     \\
          ~~~Individual beam FWHM  & 1.2\,\mathrm{arcmin}~~ \\
            ~~~Beam spacing     & 3.1\,\mathrm{arcmin}~~         \\
            \noalign{\smallskip}
            \hline
         \end{array}
     $$ 
   \end{table}
%______________________________________________________________

The basic OCRA design was based on the prototype demonstrator for the Planck Low Frequency Instrument \citep{Mandolesi} and is similar to the K-band receivers mounted on the WMAP spacecraft \citep{Jarosik}. The nominal system specification is presented in Table~1. OCRA-p has two closely spaced feeds in the secondary focus of the 32\,m telescope and thus there is very similar atmospheric emission in each beam. The basic observing mode with this radiometer involve switching continuously between the two horns and between two states of a 277$-Hz$ phase switch (0 and 180$^o$), located in one arm of the receiver. The switching frequency is much greater than the characteristic frequencies of atmospheric and gain fluctuations and those effects are dramatically reduced by taking differences between switch states and channels. The result, so-called \textit{double difference}, is the observable used for data reduction. More details and an illustration of our scan strategy can be found in \citet{Lowe}.

The full 100-beam OCRA receiver could be ideal for studying radio sources at higher frequencies, especially it is ideally suited to fast sky surveys. 

\section{Observations and data reduction}
\label{obs}
The observations were made over a period of 1.5 years, starting in December 2005. For a survey we selected 442 planetary nebulae from the \textit{Strasbourg-ESO Catalogue of Galactic Planetary Nebulae} \citep{AckerCAT} having 5\,GHz observations and declinations above $-15^o$.

A single flux density measurement consisted of two orthogonal scans with the first one made in elevation. Real-time software fitted a Gaussian profile to the observed beam shape and the measured position offset was used to update the telescope pointing corrections for the subsequent azimuthal scan. These corrections of pointing accuracy led to reliable estimate of flux density using the azimuthal scans. The first step of the data reduction was to fit a quadratic background baseline, subtract it from the data and fit double Gaussians to the azimuthal scan to ensure that the source is passing through each beam in turn. The quality of the data was strongly dependent on the weather conditions. During bad weather large asymmetries would be observed in the fitted amplitudes of the two beams. Data with differences in amplitudes exceeding 10\% were not used. To fix the flux scale, NGC~7027 was used as the calibrator for the whole survey. It is the brightest planetary nebula in the radio sky due to its proximity and youth. It is a compact nebula of 10x13.5\,arcsec \citep{Atherton} observed regularly at the VLA for 25\,years and can be taken to be unresolved when observed with the 1.2\,arcmin beam of the 32\,m telescope. Those features make NGC~7027 a good prime calibrator over a wide frequency range. Fortunately, at Toru\'n's latitude the source is circumpolar so it could be observed at any time during the survey.

On human time scales the flux evolution of NGC~7027 is slow but it is measurable. \citet{Perley} suggest that the observed smooth decline of PNe flux densities at high frequencies is due to the reduction in the number of ionising photons with time. This is caused by evolution of the central star, whose temperature is increasing and luminosity is slowly decreasing. Measurements by \citet{Zijlstra2} show that the flux density of NGC~7027 is decreasing at a rate of -7.79\,mJy/yr at 30\,GHz. Using this secular decrease in flux density, along with a flux density of 5433.8\,mJy at 2000.0 based on VLA measurements in seven bands, we calculate a flux density for 2006.0 of 5.39$\pm$0.28\,Jy. The data for all observed nebulae were scaled to this value.

The final results were corrected for gain-elevation curve of the telescope with elevation (a factor of up to 20\%, depending of elevation). The atmospheric absorption was also taken into account. The final flux density uncertainty arising from double Gaussian fitting, calibrator uncertainty and all other corrections is about 8\%.

\section{Results}
\label{res}
The final results are presented in Table~\ref{table1}. The first column contains the SESO IAU Planetary Nebula in Galactic coordinates (PNG) name lll.l$\pm$bb.b, where lll.l is Galactic longitude and bb.b is Galactic latitude. The second column contains the normal descriptive designation. Coordinates for J2000, taken from \citet{AckerCAT}, are given in columns 3 and 4. Columns 5 and 6 contain the measured flux densities at 30\,GHz and calculated uncertainties. The values marked by $\ast$ are sources with optical angular size larger that 1/4 OCRA-p beam, while those marked by $\Delta$ are corrected using adopted angular diameter which can be found only for some sources (see section~\ref{analize}). Table~\ref{table1} contains data for 93 sources. Figure~7 presents PN modeled radio spectra (see section~\ref{analize}) together with flux densities at 1.4, 5, 14.7\,GHz.\\

\begin{figure}
   \centering
   \includegraphics[width=0.5\textwidth]{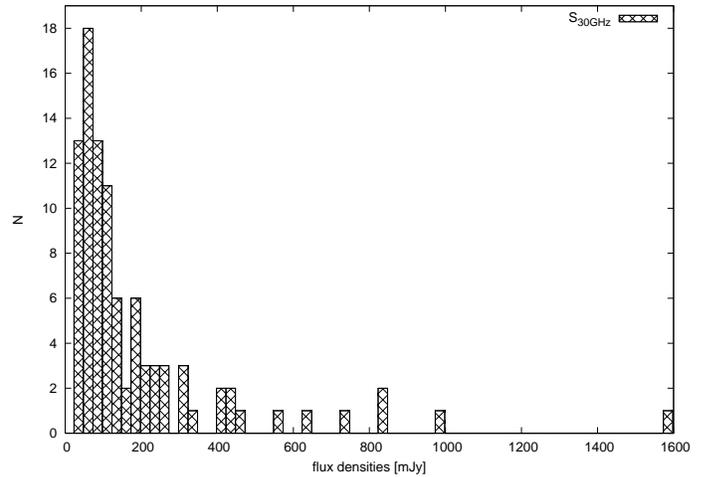}
      \caption{Histogram of the distribution of measured planetary nebulae flux densities at 30\,GHz.} 
\label{fluxS30}
   \end{figure}

The distribution of detected flux densities in presented in Figure~\ref{fluxS30}. It is seen that there is an instrumental limit on the detectability of sources around 20\,mJy. The exact value depends on the weather conditions at the time of observations.\\

\section{Analysis of flux density results}
\label{analize}
\begin{figure}
   \centering
   \includegraphics[width=0.5\textwidth]{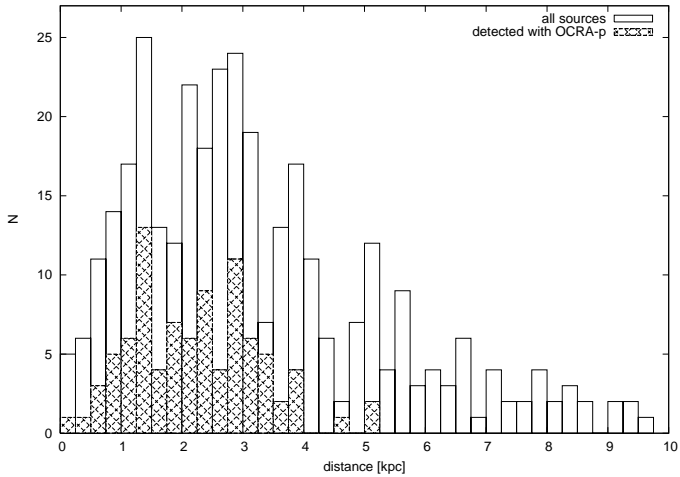}
      \caption{Histogram of the distribution of planetary nebulae distances. The plot presents distances to all planetary nebulae examined in the survey and the ones detected by OCRA-p are highlighted.}
         \label{dist}
   \end{figure}

The first result of the survey is a new catalogue of 93 Planetary Nebulae flux densities at a frequency of 30\,GHz. We have plotted in Figure~\ref{dist} a histogram for the distances of all PNe surveyed in this work, and marking those detected by OCRA-p. The data was taken from the \textit{Strasbourg-ESO Catalogue of Galactic Planetary Nebulae}. Histogram shows that we were able to measure planetary nebulae flux densities at distances out to about 5\,kpc and that the distributions of detected and non-detected PNe in our data are similar. 

\label{model}

\begin{figure*}
   \centering
   \includegraphics{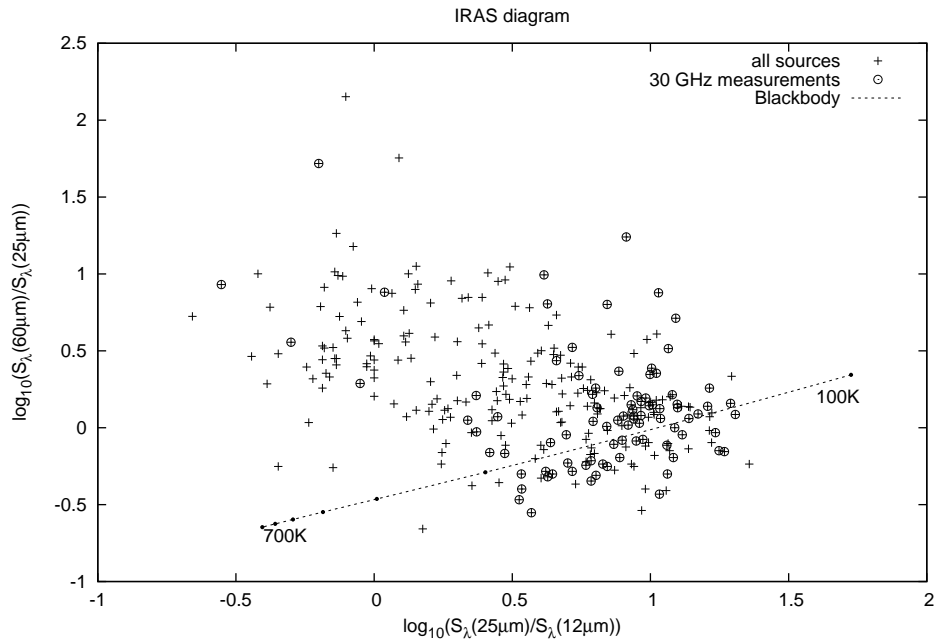}
      \caption{The distribution of planetary nebulae in the colour-colour diagram, showing all sources examined in the survey and those detected by OCRA-p. The line indicates IRAS colours for a black body distribution with temperatures from 100\,K up to 700\,K, with 100\,K steps marked by dots.}
         \label{color}
   \end{figure*}

Planetary nebulae are strong infrared sources and it is known that they occupy certain characteristic positions on the IRAS colour-colour diagram. \citet{PottaschN} made IRAS measurements of over 300 planetary nebulae in 4 bands: 12\,$\mathrm{\mu m}$, 25\,$\mathrm{\mu m}$, 60\,$\mathrm{\mu m}$ and 100\,$\mathrm{\mu m}$. We used the first three to construct a useful colour-colour diagram. Figure~\ref{color} shows this diagram for all examined sources and the ones detected with OCRA-p. It can be seen that planetary nebulae measured by OCRA are concentrated close to the black-body (120-180\,K) line in the plot. This is due to the evolution of planetary nebula. As the dust cools with time, the dust colours will redden whilst the nebulae becomes less opaque in the radio region, resulting in an increase in radio emission. Thus observed radio emission provides useful hints on planetary nebulae evolution.\\

Planetary nebula emission at 30\,GHz is thought to be due to free-free emission. We want to check whether this mechanism can fully explain the measured flux densities since there have been claims that the excess emission has been seen in planetary nebulae \citep{Casassus1}. A model of planetary nebula free-free emission was taken from \citet{Siodmiak}. These authors analysed the flux densities of 264 planetary nebulae with reliable measurements at 1.4 and 5\,GHz and known diameters (with radio diameter defined by radio contours at 10\% of the peak flux density, or the mean value of optical and radio diameter). Four simple models were proposed and examined. The most reasonable approach is to consider a model consisting of two regions with different optical depth. Each source is then modelled a dense region which fills a solid angle $\xi\Omega$ characterised by optical depth $\tau_\nu$, and a less dense region filling up $ (1-\xi) \Omega$ described by optical depth $\epsilon\tau_\nu$. With such an assumption one can model the spectra of all PNe, including the small fraction of the population which are homogeneous and symmetric. 

The observed flux density is then expressed as:
   \begin{equation}
S_\nu=\frac{2\nu^2kT_e}{c^2} \left({ \left({ 1-e^{-\tau_\nu}   } \right)   \xi\Omega +   \left({ 1-e^{-\epsilon \tau_\nu} } \right) (1-\xi) \Omega } \right),
   \end{equation}
where the optical thickness depends on the frequency $\nu$ roughly as \citep[see][]{Pottasch}: 
  \begin{equation}
\tau_\nu=\tau_0(\nu/\nu_0)^{-2.1}.
   \end{equation}
$\tau_0$ and $\nu_0$ (=5\,GHz) are the reference optical thickness and frequency. Optical thickness at 5\,GHz is found by solving the first expression with known flux density at this frequency.

The best fit values found by \citet{Siodmiak} from their population as a whole were $\xi=0.27$ and $\epsilon=0.19$. The model assumes that entire nebula has the same electron temperature, $T_e$. The values of $T_e$ were calculated based on measurements of HeII$\lambda$4686\AA~ ~taken from \citet{Tylenda} and the method described by \citet{Kaler}. With this knowledge we can count modeled flux density at 30\,GHz.

Several data sets are required to apply the model described above. HeII$\lambda$4686\AA~line intensities were taken from \citet{Tylenda}, 5\,GHz radio flux densities from \citet{AckerCAT}, 1.4\,GHz flux densities from \citet{CondonCAT} and adopted angular diameters from \citet{Siodmiak}. 5\,GHz and 30\,GHz flux density data are from single-dish measurements. This is important because, as noticed by \citet{Zijlstra3}, observed flux densities from a single dish are systematically higher than from an interferometer, which is insensitive to the extended, ionised halo emission. The data required for the model are available for 41 objects out of the 93 for which we have reliable 30\,GHz flux densities and for these sources the emission model can be tested. If source was extended in comparison to OCRA-p beam, the adopted angular diameters were used to correct measured flux densities. We assumed that PN are spherical sources with the given diameter and with constant surface brightness. This was a case of only 3 sources: NGC2438, NGC7008, NGC6781. Table~\ref{table1} contains corrected values.

\begin{figure}
   \centering
   \includegraphics[width=0.5\textwidth]{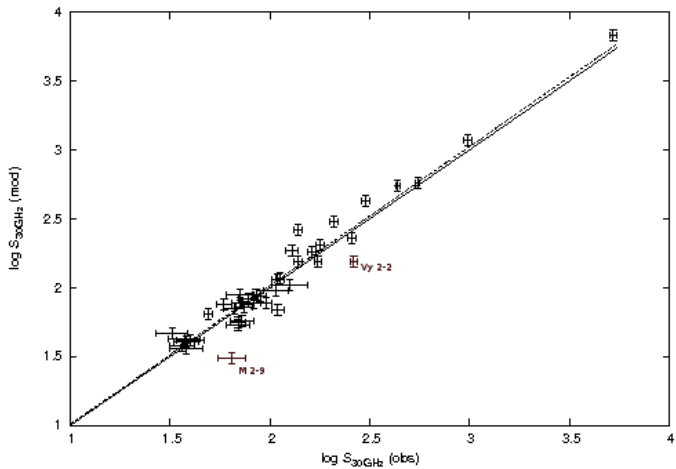}
      \caption{Modeled versus observed flux density at 30\,GHz. The solid line indicates these flux densities being equal, while the dashed line represents the best straight-line fit to relationship between $log\,S_{30GHz}\,(mod)$ and $log\,S_{30GHz}\,(obs)$.}
         \label{por}
   \end{figure}

\begin{figure}
   \centering
   \includegraphics[width=0.5\textwidth]{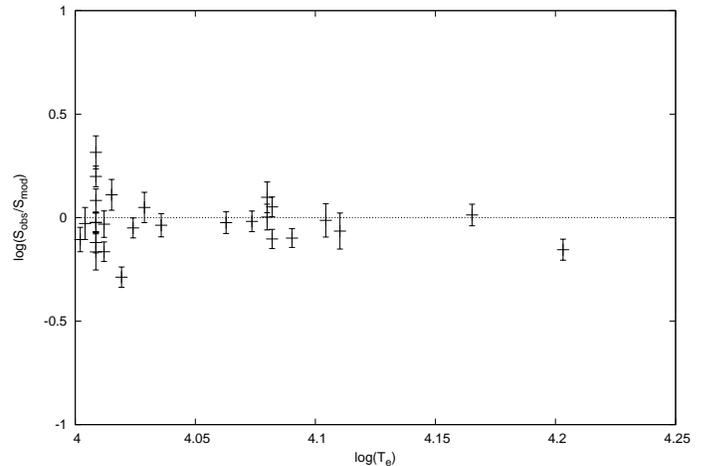}
      \caption{Observed to modeled flux density ratio at 30\,GHz versus electron temperature.}
         \label{1_4}
   \end{figure}

\begin{figure}
   \centering
   \includegraphics[width=0.5\textwidth]{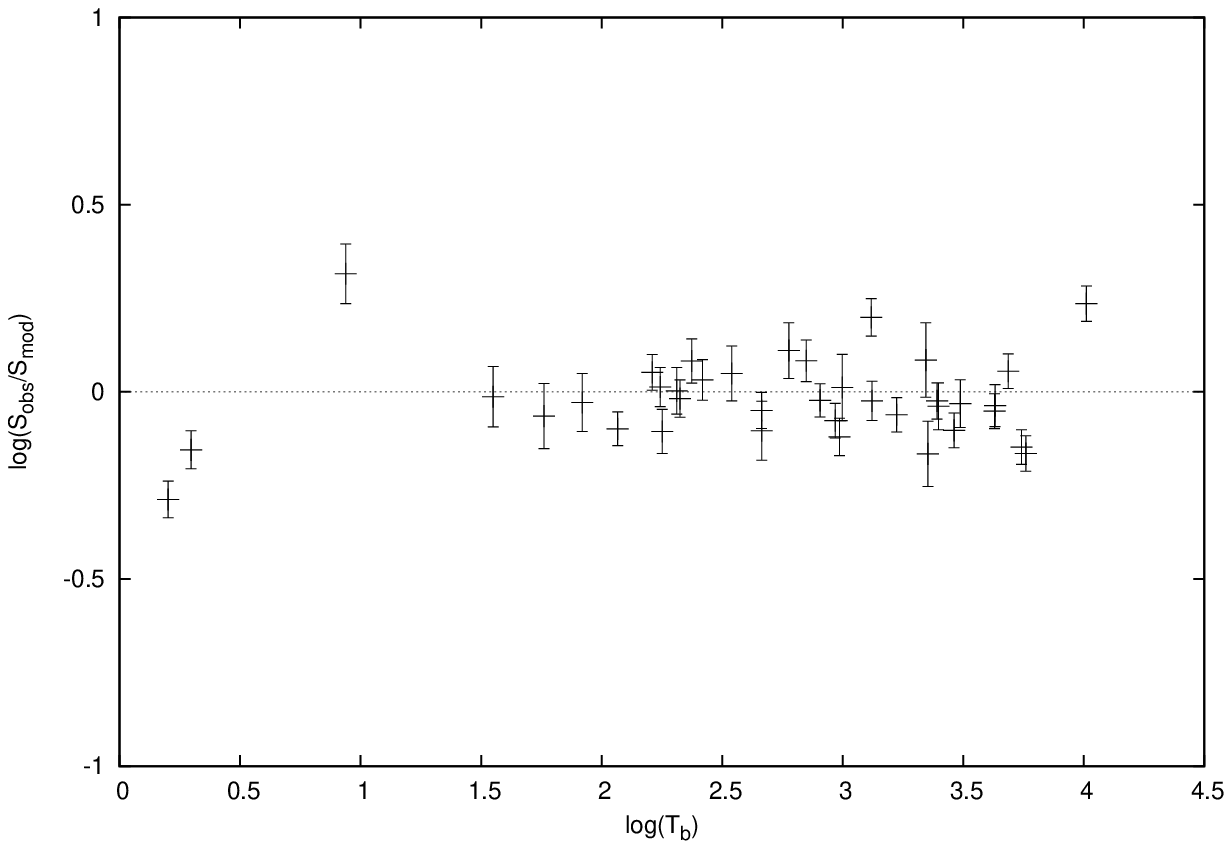}
      \caption{Observed to modeled flux density ratio at 30\,GHz versus brightness temperature.}
         \label{30}
   \end{figure}

In Figure~\ref{por} we compare the observed and modeled 30\,GHz flux densities. The model generally agrees well with the measured values. The dependence of the logarithm of the observed to modeled flux densities at 30\,GHz on electron temperature and brightness temperature (derived from flux density at 5\,GHz - $S_{5\,GHz}$ and PNe diameter - $\Theta$ using $T_b=73.87\cdot S_{5\,GHz}/{\Theta^2}$) is presented in Figures~\ref{1_4} and \ref{30}. As can be seen, the differences between observed and modeled values are small and independent of $T_e$ or $T_b$ parameters. 

Most of the data is in good agreement with the model, despite its simplicity, so that our observations can be generally explained by the free-free emission mechanism with no requirement for an additional emission mechanism. However, there are two sources, M~2-9 (010.8+18.0) and Vy2-2 (058.3-10.9), in which it can be seen that the observed flux density is significantly higher than the value expected from the model.

\textit{M~2-9} is a planetary nebula known for its additional emission at higher frequencies (see Figure~7). This emission was explained by \citet{Kwok1} and \citet{Davis} as the result of an optically thick stellar wind, which is the dominant emission source above about 20\,GHz. The high mass loss from the central star was discovered by \citet{Swings} on the basis of their analysis of its broad H$\alpha$ line. The observed 30\,GHz point shows that the stellar wind is optically thick above 30\,GHz. 

\textit{Vy~2-2} (see Figure~7) is a young and compact nebula. It is characterised by bright IR emission, a large optical depth at radio frequencies and faint emission. Its turnover frequency is much greater than 10\,GHz so the assumption of optically thin emission at 5\,GHz made by our model is not well fulfilled and therefore a disagreement with the model is not unexpected. 

Another source which should be mentioned is \textit{Hu~1-2} (086.5-08.8) (see Figure~7) where emission at 30\,GHz is a below the expected value. However a check of the other spectral measurements of this nebula suggest that the point at 5\,GHz is overestimated. 

The model was constructed to check whether any mechanism other than free-free emission is needed to explain the spectra of planetary nebulae at 30\,GHz, as suggested by \citet{Casassus2}. Such an excess was found at 10-30\,GHz in the spectral energy distribution (SED) of sources such as the dark cloud LDN~1622 \citep{Finkbeiner}, an H~II region in Perseus \citep{Watson} and the Helix nebula \citep{Casassus1}. \citet{Draine0} suggest that radiation from spinning electric dipoles created by very small grains (VSGs) or spinning dust is responsible for the observed excess. Yet in our data the effect is not observed. This conclusion is in agreement with new observations of planetary nebulae at 43\,GHz \citep{Umana}, whose measurements are mostly in agreement within errors with those presented in this paper (with exception for PN IC2165 (221.3-12.3)).

\section{Conclusions}
We surveyed 442 planetary nebulae from 5-GHz catalogue of \citet{AckerCAT} with declinations above $-15^o$. 93 sources were detected above 23\,mJy and their flux densities at 30\,GHz were measured and presented in this paper. The model for free-free emission described by \citet{Siodmiak} was used to predict 30\,GHz flux densities based on the 5\,GHz observations and the known diameters. These calculated values are in good agreement with the observed flux densities. No additional emission mechanism is needed to explain the observed spectra.
\label{con}

\begin{acknowledgements}We thank Albert Zijlstra and Krzysztof Gesicki for helpful discussions and comments. We are grateful for support from Royal Society Paul Instrument Fund and for funds from Ministry of Science in Poland (grant number N N203 390434). M.Peel acknowledges the support of an STFC studentship.
\end{acknowledgements}

\addtocounter{table}{1}
\addtocounter{table}{2}

\bibliographystyle{aa}

\longtab{2}{
\begin{longtable}{llllrrr}
\caption{\label{table1} 30\,GHz survey.; col. (1) gives the SESO IAU Planetary Nebula in Galactic coordinates (PNG) name lll.l$\pm$bb.b, where lll.l is Galactic longitude and bb.b is Galactic latitude. The values marked by $\ast$ are sources with optical angular size larger that 1/4 OCRA-p beam, while those marked by $\Delta$ are corrected using adopted angular diameter (see section~\ref{analize}).; col. (2) gives the normal descriptive designation, cols. (3) and (4) give coordinates for J2000, taken from \citet{AckerCAT}; cols. (5) and (6) give measured flux densities at 30\,GHz and calculated uncertainties.}\\
\hline\hline
PNG & Name & R.A. (J2000) & Declination (J2000) & $S_{30GHz}$ (mJy) & $\Delta S_{30GHz}$ (mJy) \\ 
(1)&(2)&(3)&(4)&(5)&(6)\\
\hline
\endfirsthead
\caption{continued.}\\
\hline\hline
PNG & Name & R.A. (J2000) & Declination (J2000) & $S_{30GHz}$ (mJy) & $\Delta S_{30GHz}$ (mJy) \\   
(1)&(2)&(3)&(4)&(5)&(6)\\
\hline
\endhead

\hline
\endfoot
 010.8+18.0&  M 2-9       & 17 05 37.952 &  -10 08 34.58&      63.8&      9.9\\
 015.9+03.3&  M 1-39      & 18 07 30.699 &  -13 28 47.61&        70&       11\\
 019.4-05.3&  M 1-61      & 18 45 55.072 &  -14 27 37.93&       127&       27\\
 019.7+03.2&  M 3-25      & 18 15 16.967 &  -10 10 09.47&      76.8&      8.0\\
 020.9-01.1&  M 1-51      & 18 33 28.943 &  -11 07 26.41&     229.9&      7.4\\
 023.9-02.3&  M 1-59      & 18 43 20.178 &  -09 04 48.64&       107&       15\\
 025.3+40.8&  IC 4593     & 16 11 44.544 &  +12 04 17.06&      66.8&      4.8\\
 025.8-17.9&  NGC 6818    & 19 43 57.844 &  -14 09 11.91&       208&       30\\
 027.6+04.2&  M 2-43      & 18 26 40.048 &  -02 42 57.63&       259&       10\\
 027.7+00.7&  M 2-45      & 18 39 21.837 &  -04 19 50.90&       158&       17\\
 031.0+04.1&  K 3-6       & 18 33 17.490 &  +00 11 47.19&      98.6&      8.3\\
 032.7-02.0&  M 1-66      & 18 58 26.247 &  -01 03 45.70&      68.7&      9.6\\
 034.0+02.2&  K 3-13      & 18 45 24.582 &  +02 01 23.85&      91.3&      7.1\\
 034.6+11.8&  NGC 6572    & 18 12 06.365 &  +06 51 13.01&       987&       38\\
 035.1-00.7&  Ap 2-1      & 18 58 10.459 &  +01 36 57.15&       241&       13\\
 037.7-34.5&  NGC 7009    & 21 04 10.877 &  -11 21 48.25&       470&       21\\
 037.8-06.3&  NGC 6790    & 19 22 56.965 &  +01 30 46.45&       303&       12\\
 038.2+12.0&  Cn 3-1      & 18 17 34.104 &  +10 09 03.34&      68.6&      6.4\\
 039.8+02.1&  K 3-17      & 18 56 18.171 &  +07 07 26.31&       303&      9.0\\
 040.3-00.4&  A 53        & 19 06 45.910 &  +06 23 52.47&      54.9&      6.5\\
 041.8-02.9$^\Delta$&  NGC 6781    & 19 18 28.085 &  +06 32 19.29&     264.1&      7.1\\
 043.1+37.7&  NGC 6210    & 16 44 29.491 &  +23 47 59.68&     225.7&      8.1\\
 045.4-02.7&  Vy 2-2      & 19 24 22.229 &  +09 53 56.66&       266&       12\\
 045.7-04.5&  NGC 6804    & 19 31 35.175 &  +09 13 32.01&      78.1&      6.6\\
 046.4-04.1&  NGC 6803    & 19 31 16.490 &  +10 03 21.88&      83.8&      6.3\\
 047.0+42.4*&  A 39        & 16 27 33.737 &  +27 54 33.44&       264&       18\\
 048.0-02.3&  PB 10       & 19 28 14.391 &  +12 19 36.17&      37.1&      6.5\\
 048.1+01.1&  K 3-29      & 19 15 30.561 &  +14 03 49.83&      71.2&      5.8\\
 048.5+04.2&  K 4-16      & 19 04 51.471 &  +15 47 37.83&       107&       11\\
 048.7+01.9&  He 2-429    & 19 13 38.422 &  +14 59 19.21&      62.1&      6.1\\
 050.1+03.3*&  M 1-67      & 19 11 30.877 &  +16 51 38.16&     102.8&      6.1\\
 051.0+03.0&  He 2-430    & 19 14 04.197 &  +17 31 32.92&      37.7&      6.7\\
 051.4+09.6&  Hu 2-1      & 18 49 47.565 &  +20 50 39.34&     111.7&      5.5\\
 051.9-03.8&  M 1-73      & 19 41 09.323 &  +14 56 59.37&      36.5&      4.8\\
 054.1-12.1&  NGC 6891    & 20 15 08.838 &  +12 42 15.63&      97.2&      7.3\\
 055.5-00.5&  M 1-71      & 19 36 26.927 &  +19 42 23.99&     178.9&      6.3\\
 056.0+02.0&  K 3-35      & 19 27 44.036 &  +21 30 03.83&      60.1&      6.4\\
 057.9-01.5&  He 2-447    & 19 45 22.164 &  +21 20 04.03&      65.5&      7.1\\
 058.3-10.9&  IC 4997     & 20 20 08.741 &  +16 43 53.71&     108.9&      6.4\\
 060.1-07.7&  NGC 6886    & 20 12 42.813 &  +19 59 22.65&      73.5&      7.6\\
 060.8-03.6*&  NGC 6853    & 19 59 36.340 &  +22 43 16.09&        95&       11\\
 061.4-09.5*&  NGC 6905    & 20 22 22.940 &  +20 06 16.80&      38.8&      4.5\\
 063.1+13.9*&  NGC 6720    & 18 53 35.079 &  +33 01 45.03&     189.7&      7.0\\
 064.7+05.0&  BD+30 3639  & 19 34 45.232 &  +30 30 58.94&       551&       13\\
 064.9-02.1&  K 3-53      & 20 03 22.475 &  +27 00 54.73&      73.8&      5.3\\
 067.9-00.2&  K 3-52      & 20 03 11.435 &  +30 32 34.15&      90.2&      9.9\\
 068.3-02.7&  He 2-459    & 20 13 57.898 &  +29 33 55.94&     102.5&      6.9\\
 069.7-00.0&  K 3-55      & 20 06 56.210 &  +32 16 33.60&      76.9&      7.0\\
 071.6-02.3&  M 3-35      & 20 21 03.769 &  +32 29 23.86&     161.8&      6.4\\
 074.5+02.1&  NGC 6881    & 20 10 52.464 &  +37 24 41.18&     109.1&      7.4\\
 075.7+35.8&  Sa 4-1      & 17 13 50.35  &  +49 16 11.0 &     183.8&      7.6\\
 078.3-02.7&  K 4-53      & 20 42 16.744 &  +37 40 25.58&      44.4&      6.3\\
 082.1+07.0&  NGC 6884    & 20 10 23.66  &  +46 27 39.8 &     139.3&      6.8\\
 083.5+12.7&  NGC 6826    & 19 44 48.150 &  +50 31 30.26&     303.5&      8.2\\
 086.5-08.8&  Hu 1-2      & 21 33 08.349 &  +39 38 09.57&      89.2&      5.9\\
 089.0+00.3&  NGC 7026    & 21 06 18.209 &  +47 51 05.35&     208.2&      9.5\\
 089.8-00.6&  Sh 1-89     & 21 14 07.628 &  +47 46 22.17&       831&       45\\
 089.8-05.1&  IC 5117     & 21 32 31.027 &  +44 35 48.53&     208.6&      9.1\\
 093.4+05.4$^\Delta$&  NGC 7008    & 21 00 32.503 &  +54 32 36.18&     178.1&      8.1\\
 093.5+01.4&  M 1-78      & 21 20 44.809 &  +51 53 27.88&       846&       18\\
 095.2+00.7&  K 3-62      & 21 31 50.203 &  +52 33 51.64&     101.3&      7.8\\
 096.4+29.9&  NGC 6543    & 17 58 33.423 &  +66 37 59.52&       630&       15\\
 097.5+03.1*&  A 77        & 21 32 10.238 &  +55 52 42.67&     409.3&      9.8\\
 100.0-08.7&  Me 2-2      & 22 31 43.686 &  +47 48 03.96&      32.1&      5.6\\
 100.6-05.4&  IC 5217     & 22 23 55.725 &  +50 58 00.43&      38.4&      5.7\\
 104.1+01.0&  Bl 2-1      & 22 20 16.638 &  +58 14 16.59&        45&      4.6\\
 106.5-17.6&  NGC 7662    & 23 25 53.6   &  +42 32 06   &       435&       13\\
 107.8+02.3&  NGC 7354    & 22 40 19.940 &  +61 17 08.10&       439&       13\\
 110.1+01.9&  PM 1-339    & 22 58 54.857 &  +61 57 57.77&      31.4&      4.1\\
 111.8-02.8&  Hb 12       & 23 26 14.814 &  +58 10 54.65&     398.7&      9.4\\
 120.0+09.8*&  NGC 40      & 00 13 01.015 &  +72 31 19.09&     331.8&      8.3\\
 123.6+34.5&  IC 3568     & 12 33 06.871 &  +82 33 48.95&      58.3&      5.4\\
 130.2+01.3&  IC 1747     & 01 57 35.896 &  +63 19 19.36&      67.4&      3.6\\
 130.9-10.5*&  NGC 650-51  & 01 42 19.948 &  +51 34 31.15&      62.9&      3.8\\
 138.8+02.8&  IC 289      & 03 10 19.273 &  +61 19 00.91&      98.8&      4.0\\
 144.5+06.5*&  NGC 1501    & 04 06 59.190 &  +60 55 14.34&     136.2&      3.8\\
 147.4-02.3&  M 1-4       & 03 41 43.428 &  +52 17 00.28&      59.4&      5.5\\
 148.4+57.0*&  NGC 3587    & 11 14 47.734 &  +55 01 08.50&      23.1&      3.6\\
 161.2-14.8&  IC 2003     & 03 56 21.984 &  +33 52 30.59&        40&      6.3\\
 165.5-15.2*&  NGC 1514    & 04 09 16.984 &  +30 46 33.47&      59.6&      3.2\\
 166.1+10.4&  IC 2149     & 05 56 23.908 &  +46 06 17.32&     117.2&      3.5\\
 166.4-06.5&  CRL 618     & 04 42 53.64  &  +36 06 53.4 &       731&       20\\
 173.7+02.7&  PP 40       & 05 40 53.333 &  +35 42 22.29&       173&      4.9\\
 184.0-02.1&  M 1-5       & 05 46 50.000 &  +24 22 02.32&      49.3&      2.8\\
 189.1+19.8*&  NGC 2371-72 & 07 25 34.720 &  +29 29 25.63&        25&      3.9\\
 194.2+02.5&  J 900       & 06 25 57.275 &  +17 47 27.19&      86.1&      4.8\\
 196.6-10.9&  NGC 2022    & 05 42 06.229 &  +09 05 10.75&      70.1&      3.8\\
 197.8+17.3&  NGC 2392    & 07 29 10.767 &  +20 54 42.49&     186.6&      5.8\\
 206.4-40.5&  NGC 1535    & 04 14 15.762 &  -12 44 22.03&     125.4&      9.8\\
 211.2-03.5&  M 1-6       & 06 35 45.126 &  -00 05 37.36&        95&      7.6\\
 215.2-24.2&  IC 418      & 05 27 28.204 &  -12 41 50.26&      1581&       50\\
 221.3-12.3&  IC 2165     & 06 21 42.775 &  -12 59 13.96&     173.6&      7.9\\
 231.8+04.1$^\Delta$& NGC 2438     & 07 41 51.426 &  -14 43 54.88&        89&       10\\
\end{longtable}
}

\renewcommand*{\tablename}{Fig.}
\longtab{7}{
\begin{longtable}{lll}
\caption{\label{table2} Radio spectra of PN with flux densities at 1.4, 5, 14.7\,GHz from the literature \citep[see][]{CondonCAT,AckerCAT,Aaquist,Milne} and 30\,GHz (OCRA-p) points from Table~\ref{table1}, cols. (5) and (6).}\\
\endfirsthead
\caption{continued.}\\
\endhead
\endfoot
   \includegraphics[width=0.3\textwidth]{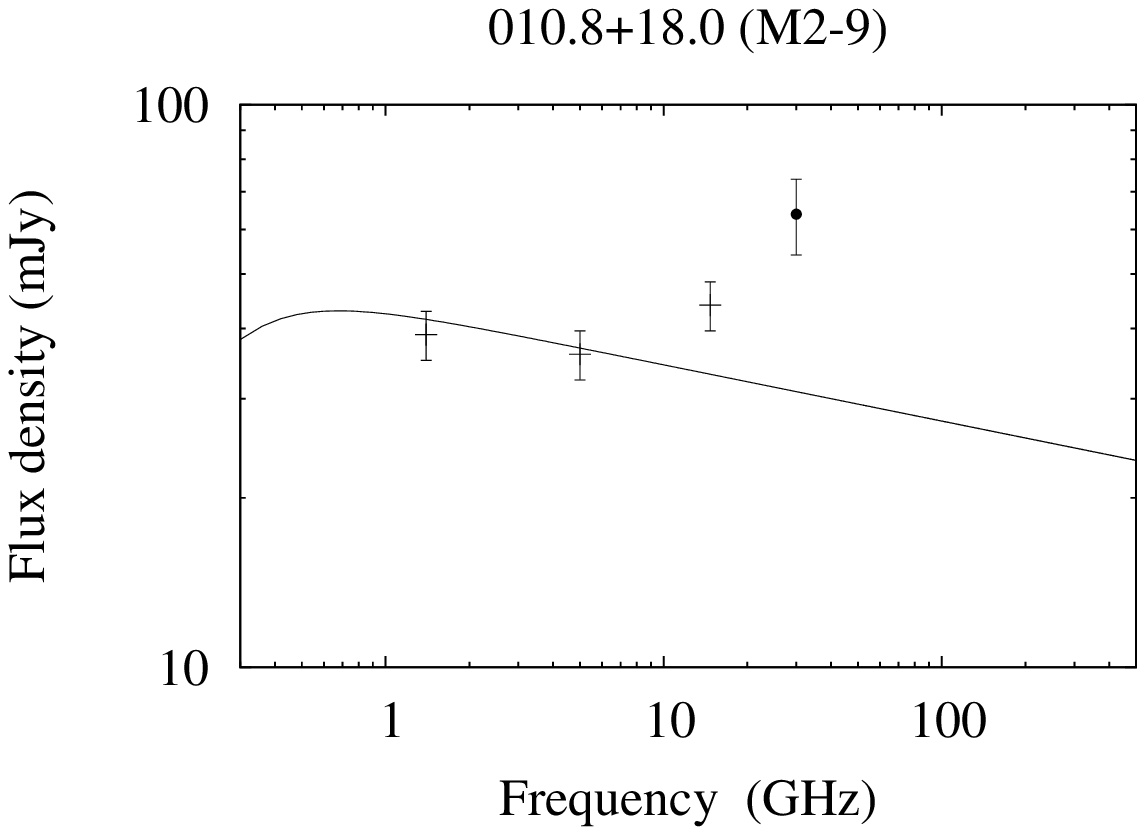}
&
   \includegraphics[width=0.3\textwidth]{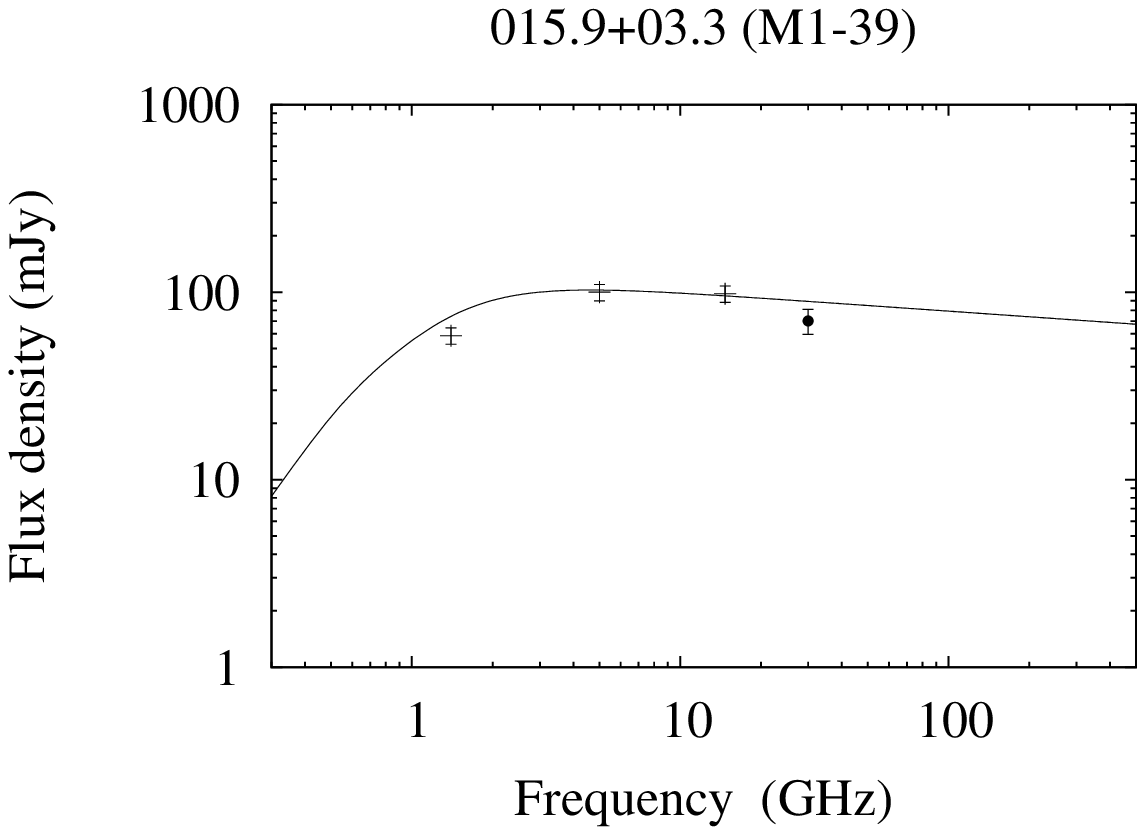}
&
   \includegraphics[width=0.3\textwidth]{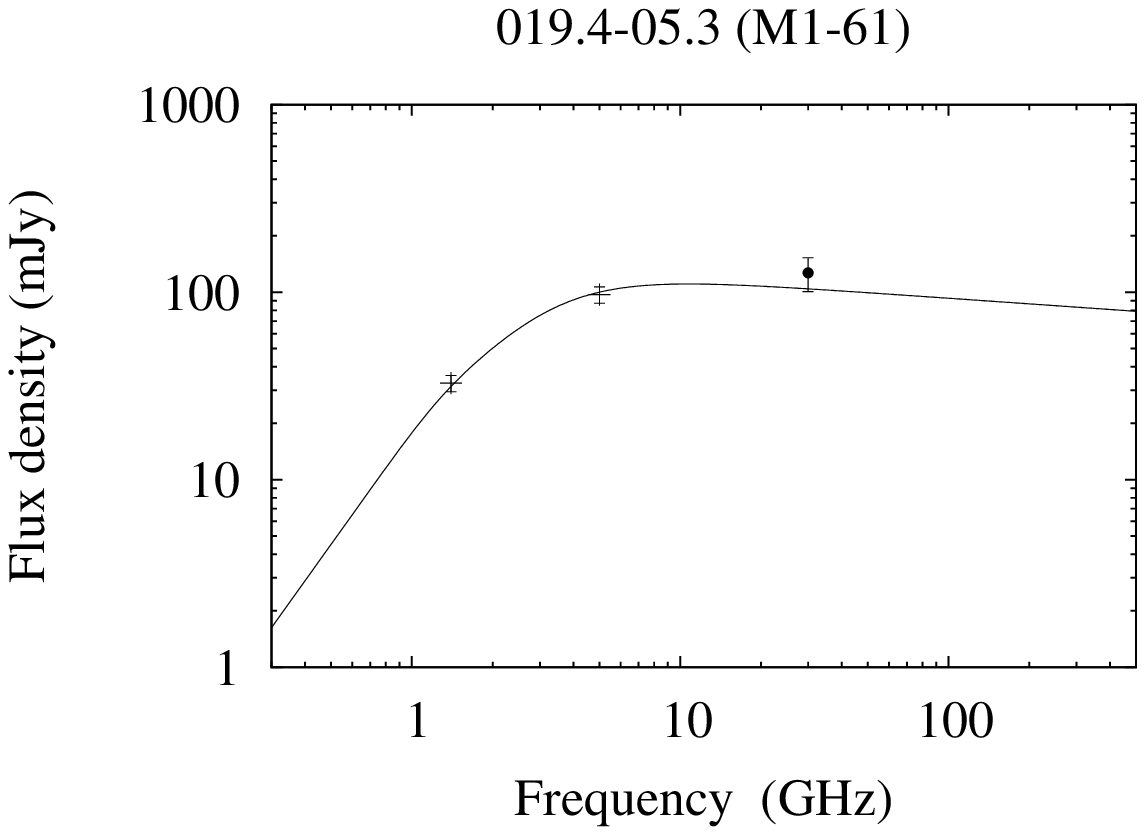}
\\
   \includegraphics[width=0.3\textwidth]{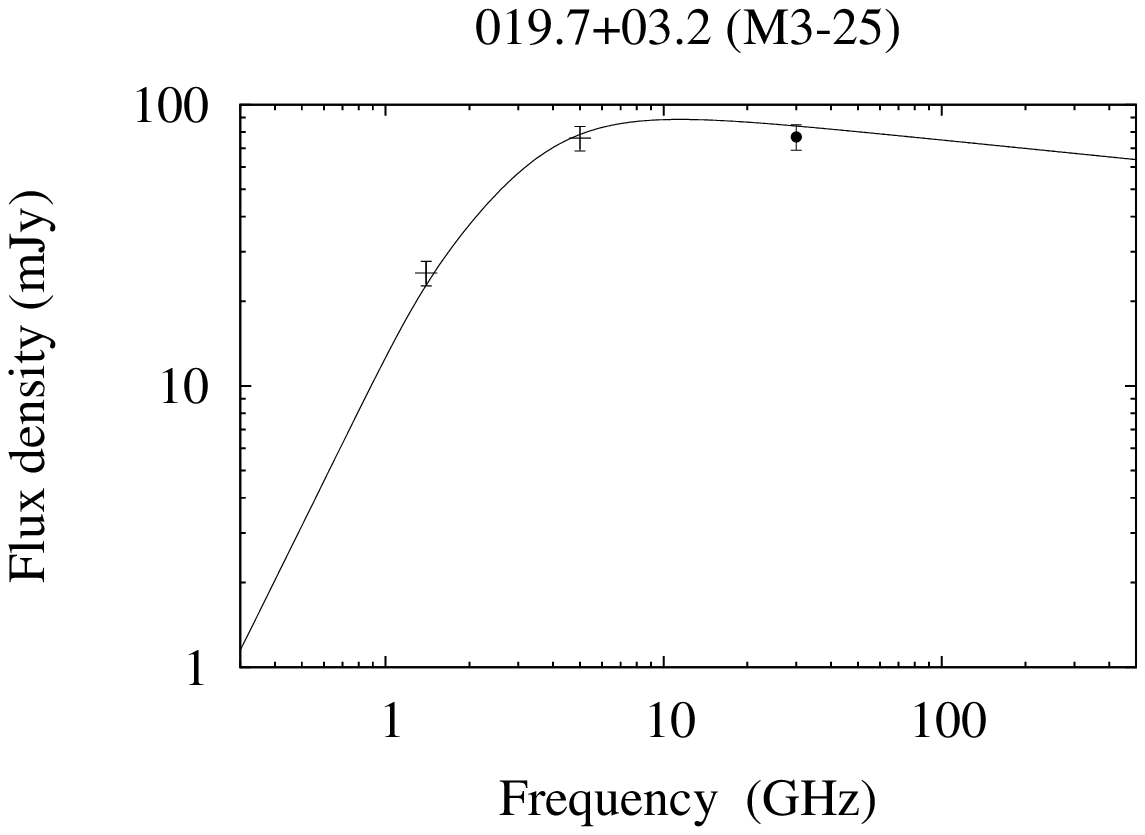}
&
   \includegraphics[width=0.3\textwidth]{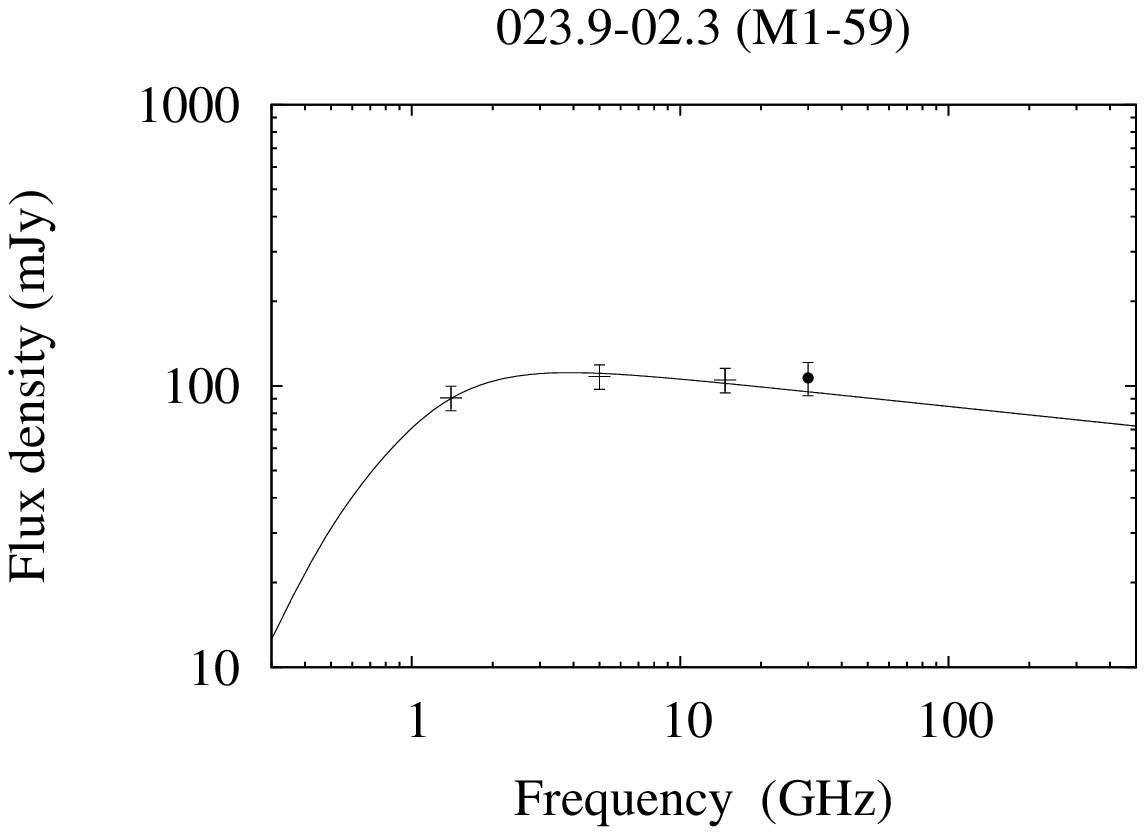}
&
   \includegraphics[width=0.3\textwidth]{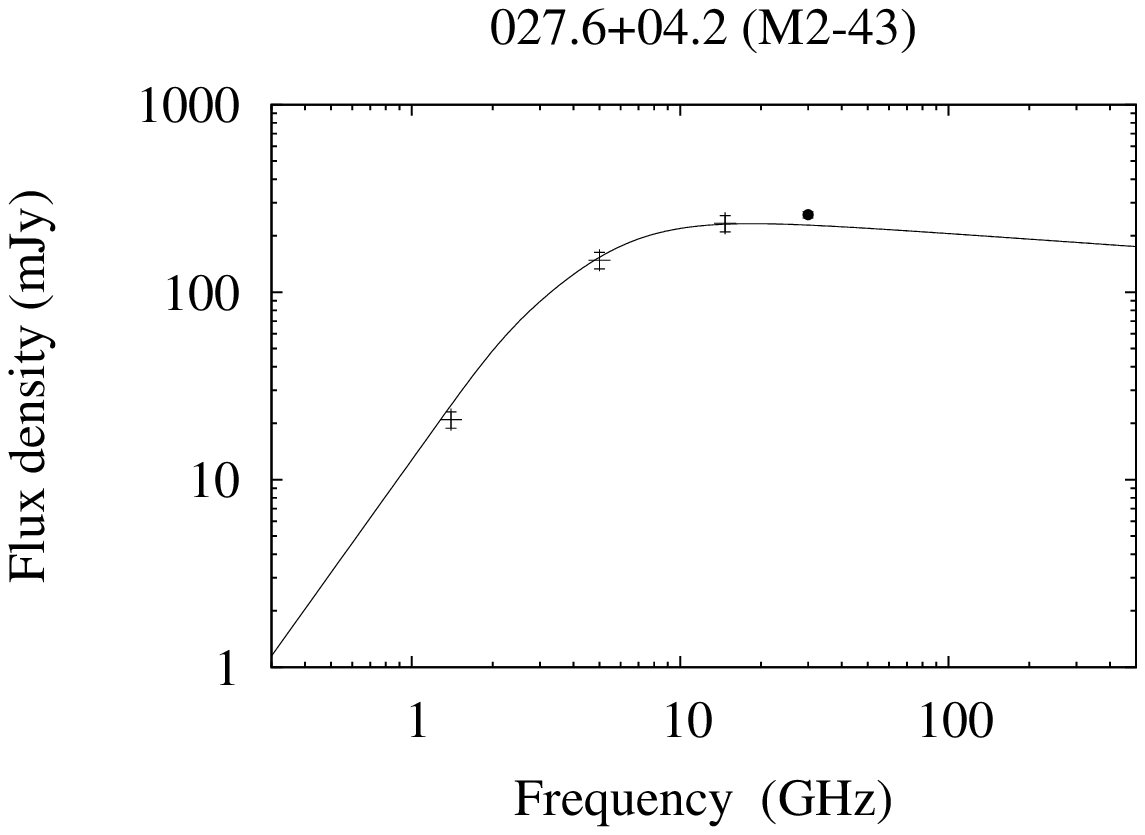}
\\
   \includegraphics[width=0.3\textwidth]{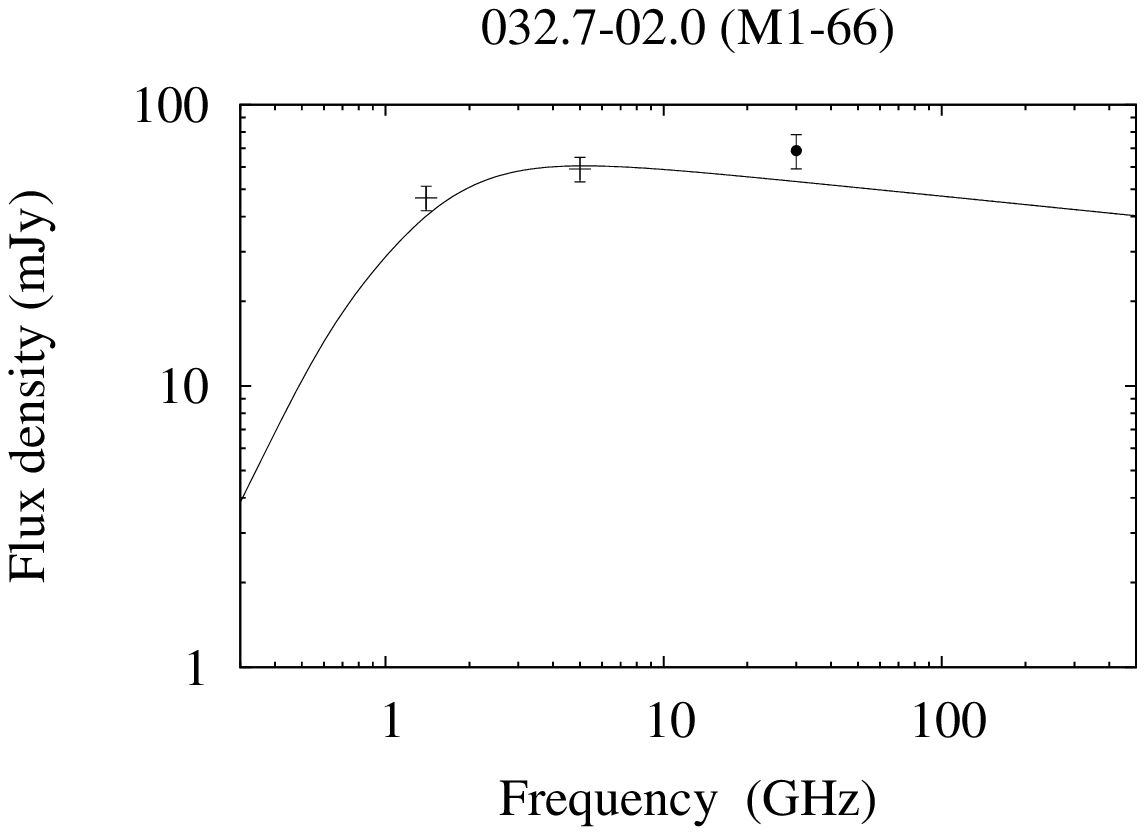}
&
   \includegraphics[width=0.3\textwidth]{032.7-02.0Q.eps}
&
   \includegraphics[width=0.3\textwidth]{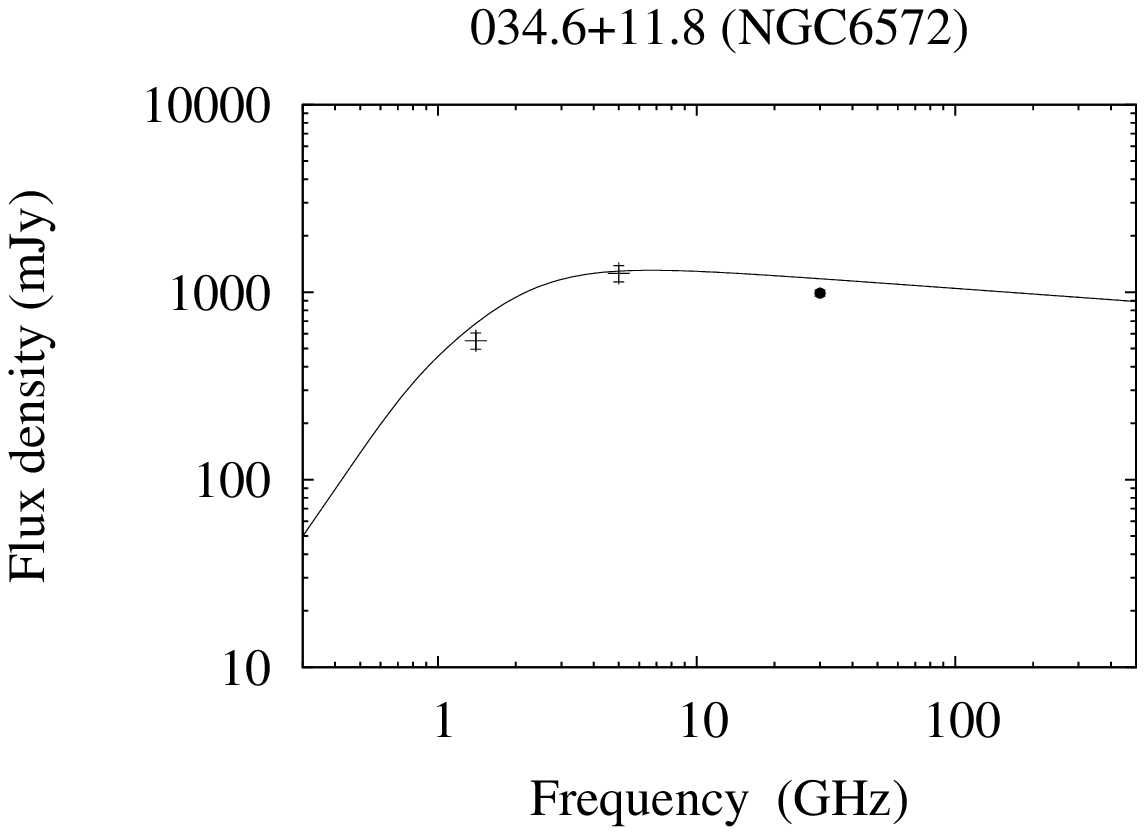}
\\
   \includegraphics[width=0.3\textwidth]{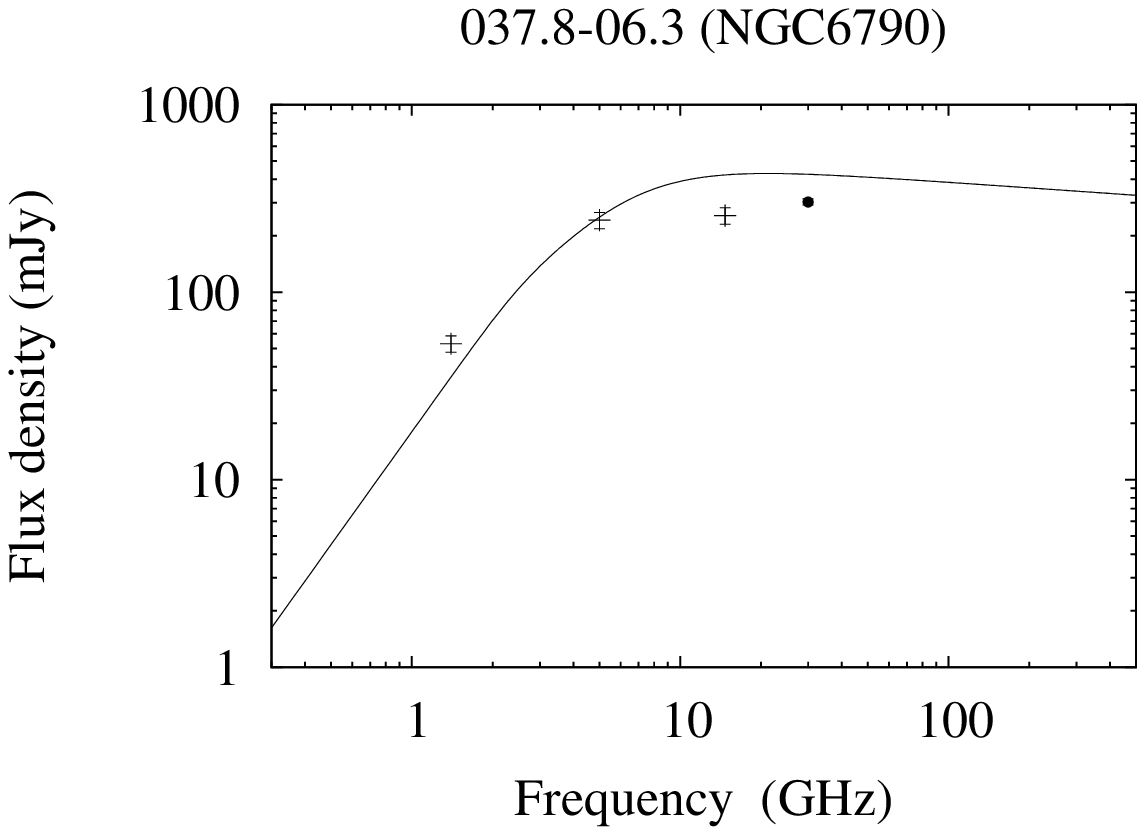} 
&
   \includegraphics[width=0.3\textwidth]{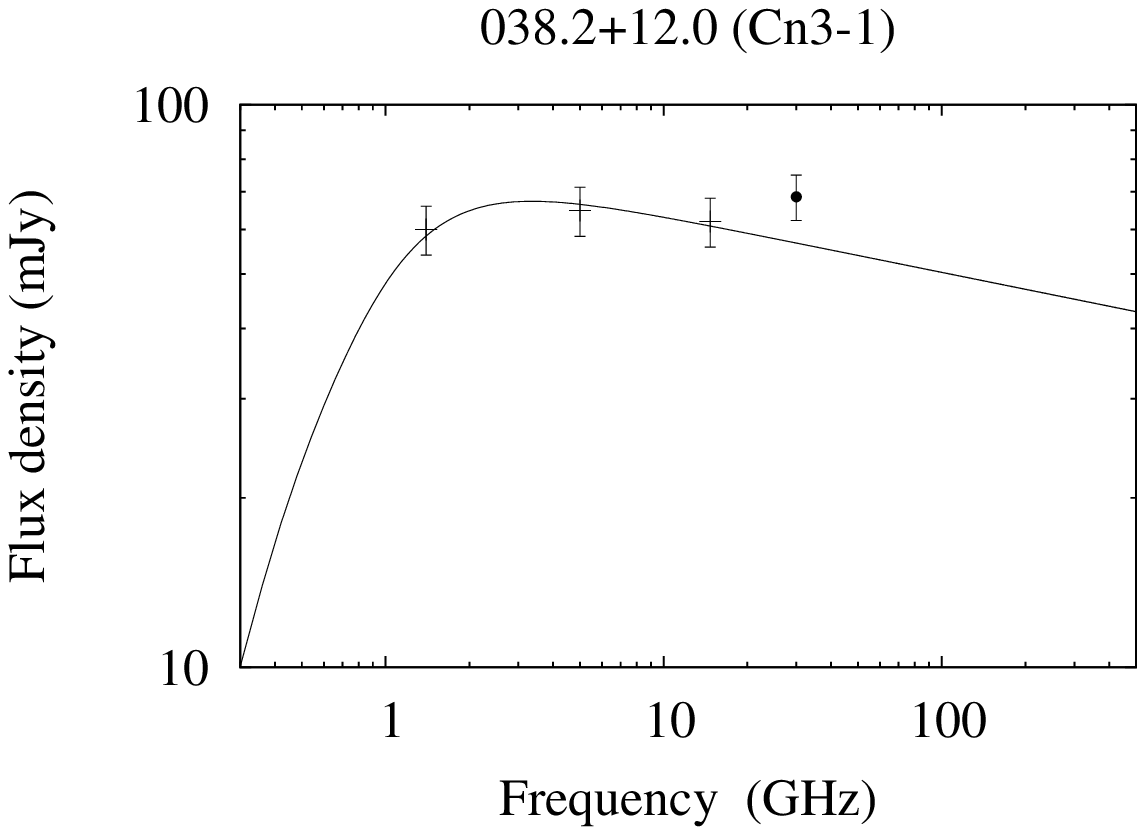}
&
   \includegraphics[width=0.3\textwidth]{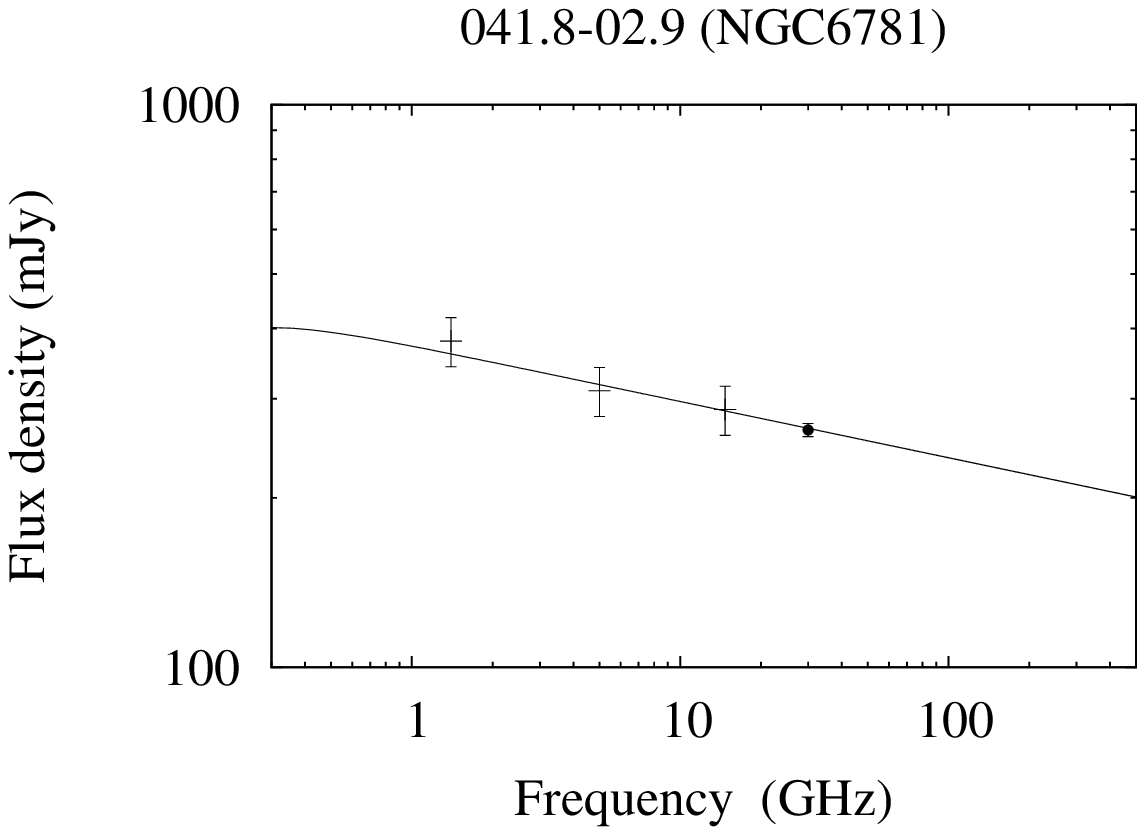}
\\
   \includegraphics[width=0.3\textwidth]{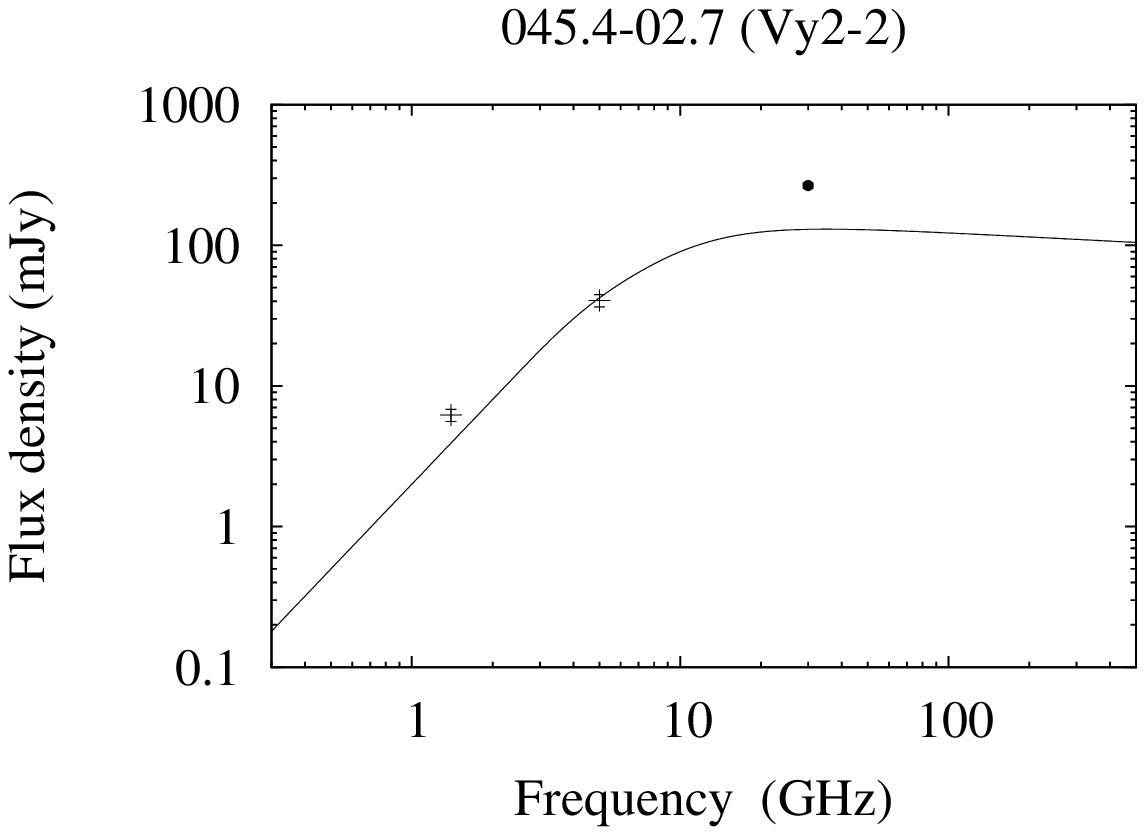} 
&
   \includegraphics[width=0.3\textwidth]{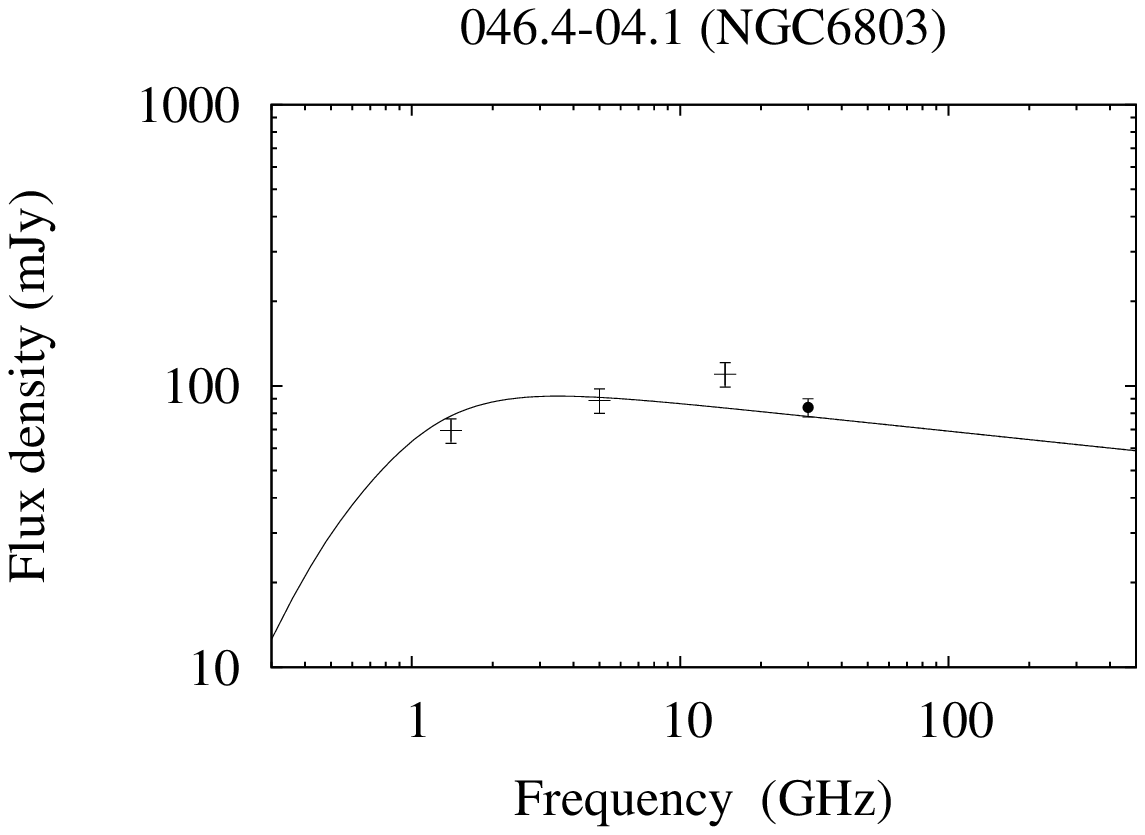} 
&
   \includegraphics[width=0.3\textwidth]{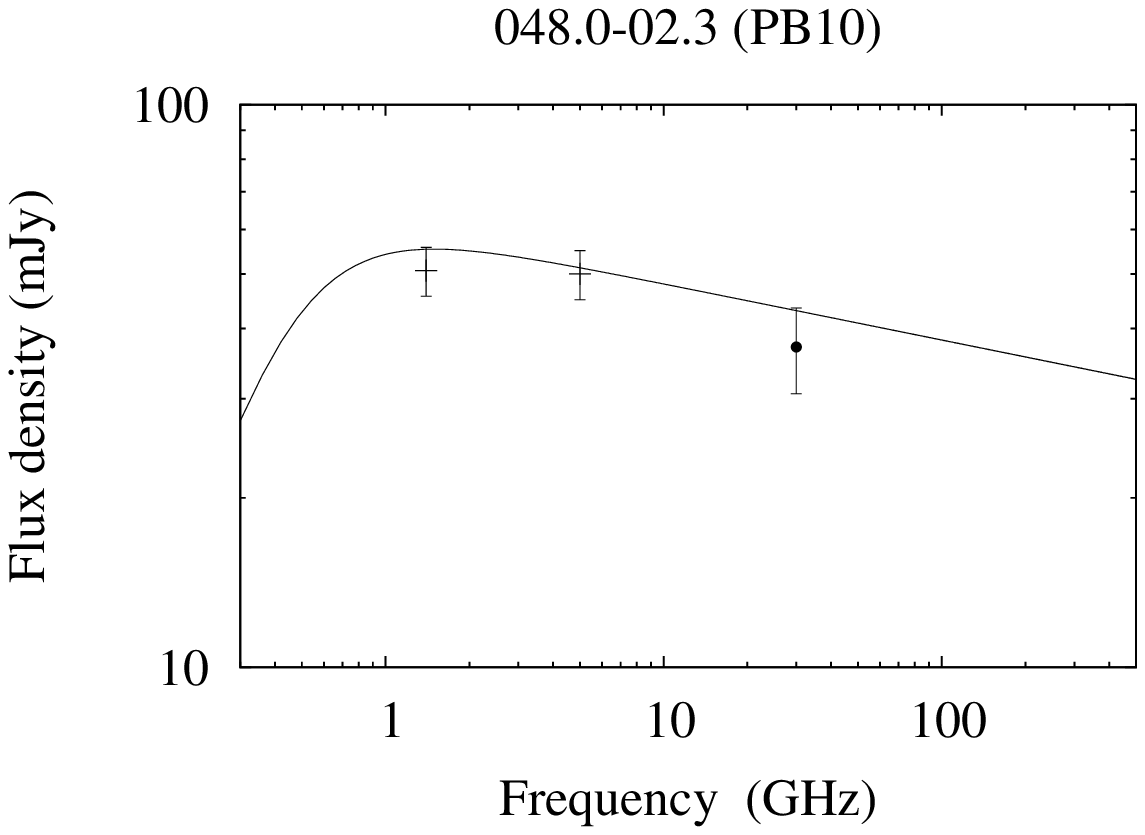} 
\\
   \includegraphics[width=0.3\textwidth]{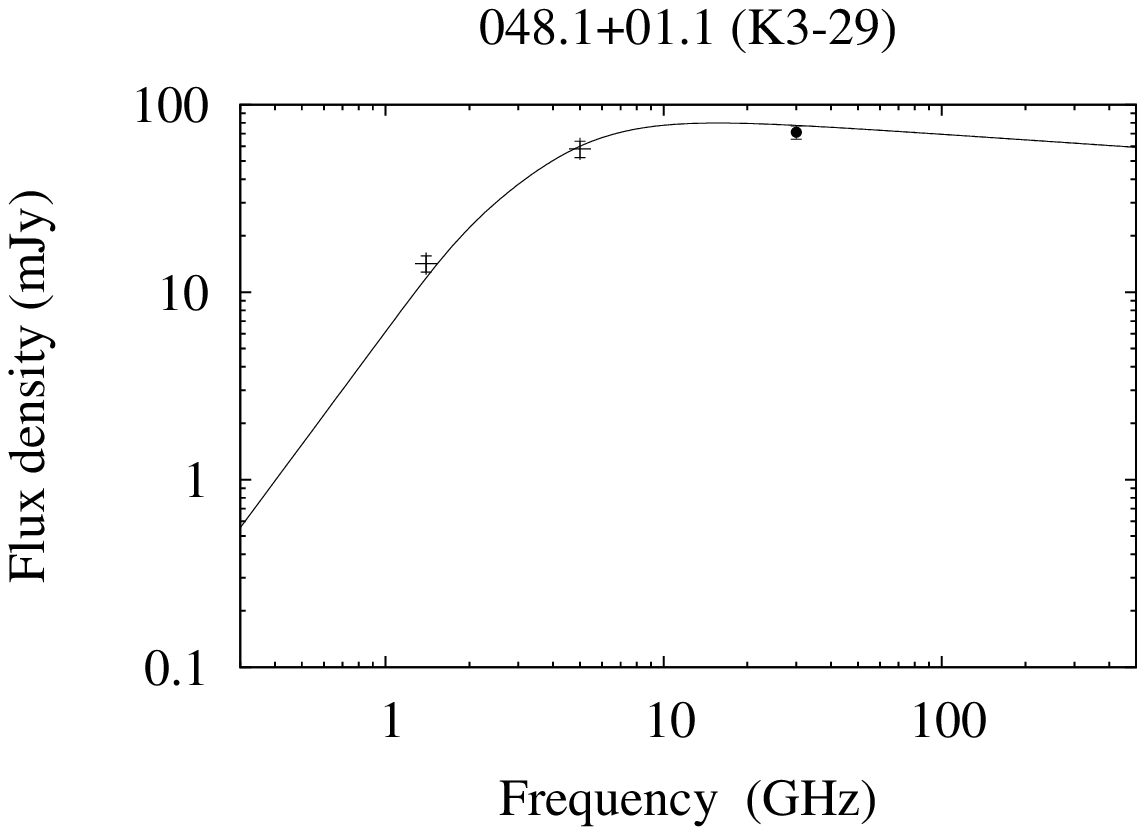}
&
   \includegraphics[width=0.3\textwidth]{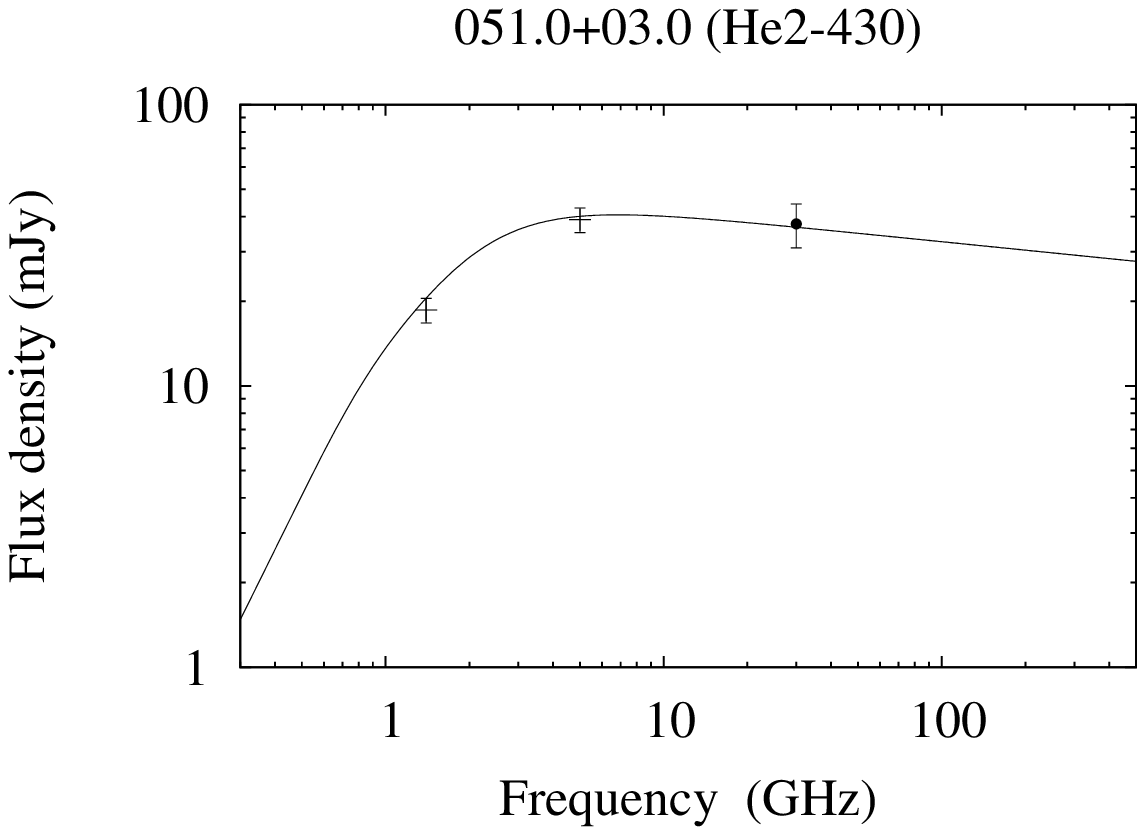}
&
   \includegraphics[width=0.3\textwidth]{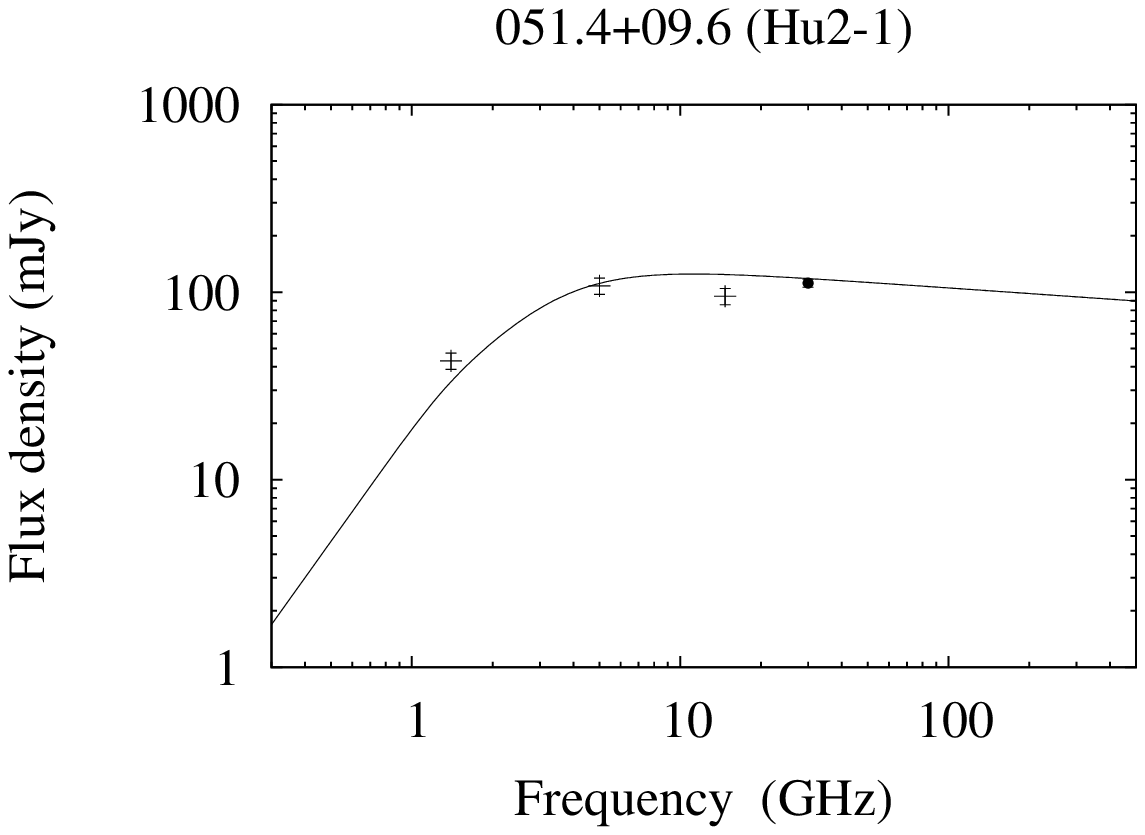}      
\\
   \includegraphics[width=0.3\textwidth]{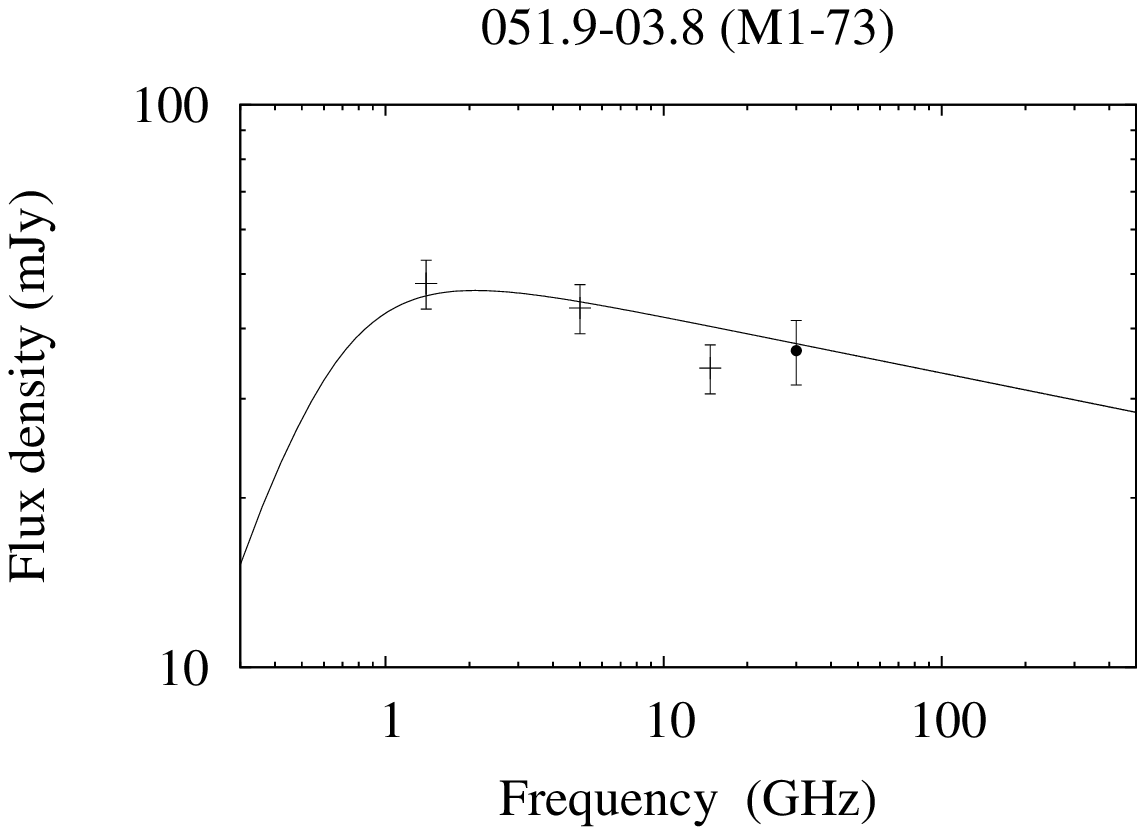}      
&
   \includegraphics[width=0.3\textwidth]{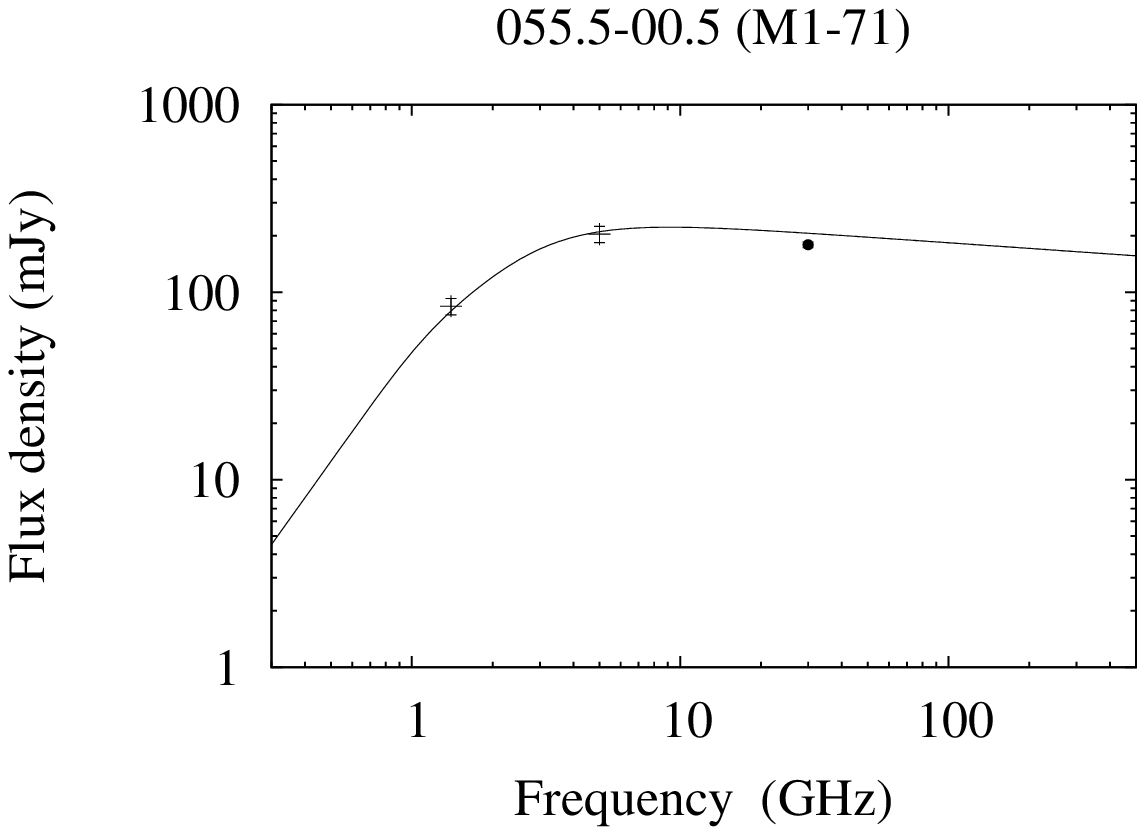}      
&
   \includegraphics[width=0.3\textwidth]{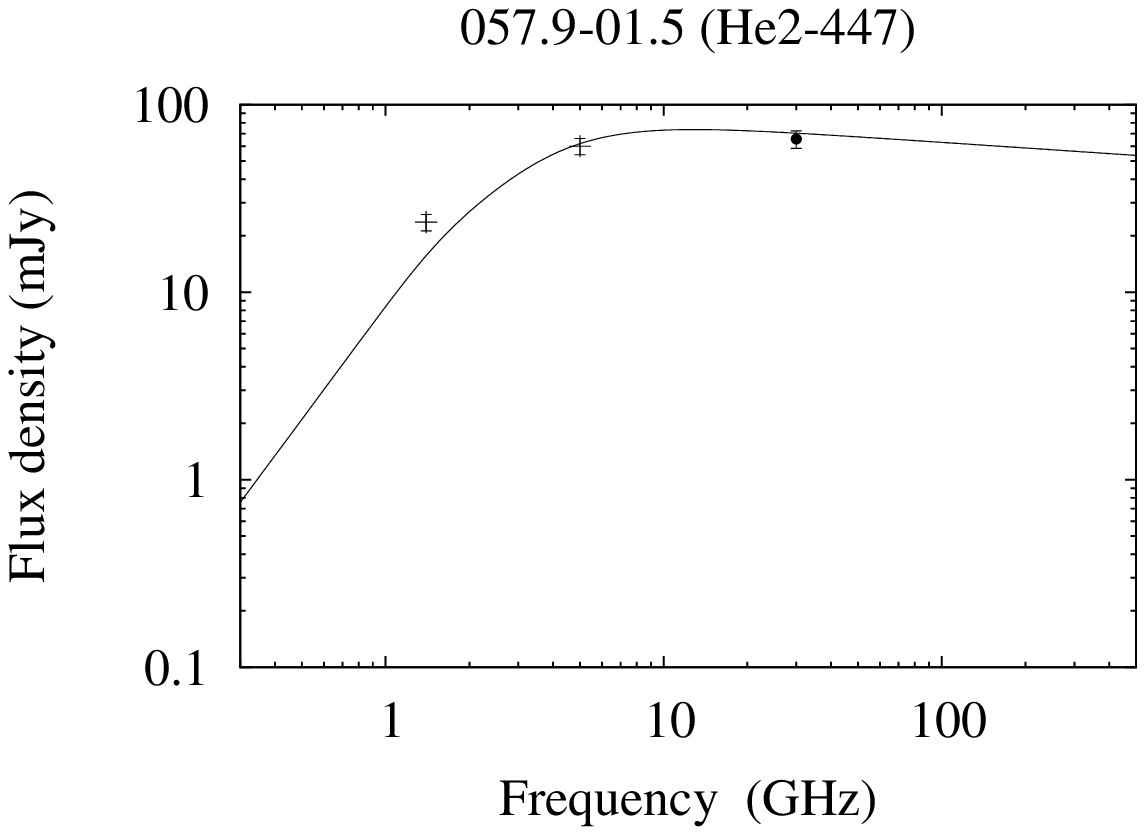}      
\\
   \includegraphics[width=0.3\textwidth]{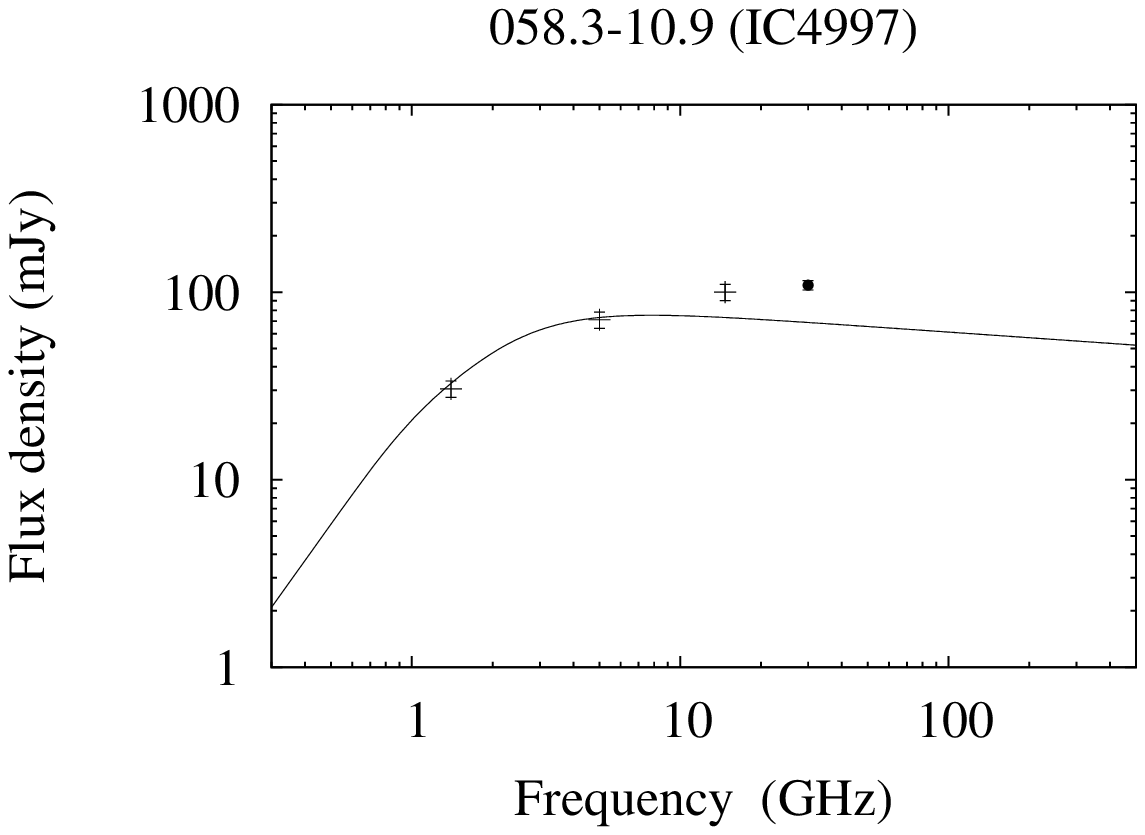}      
&
   \includegraphics[width=0.3\textwidth]{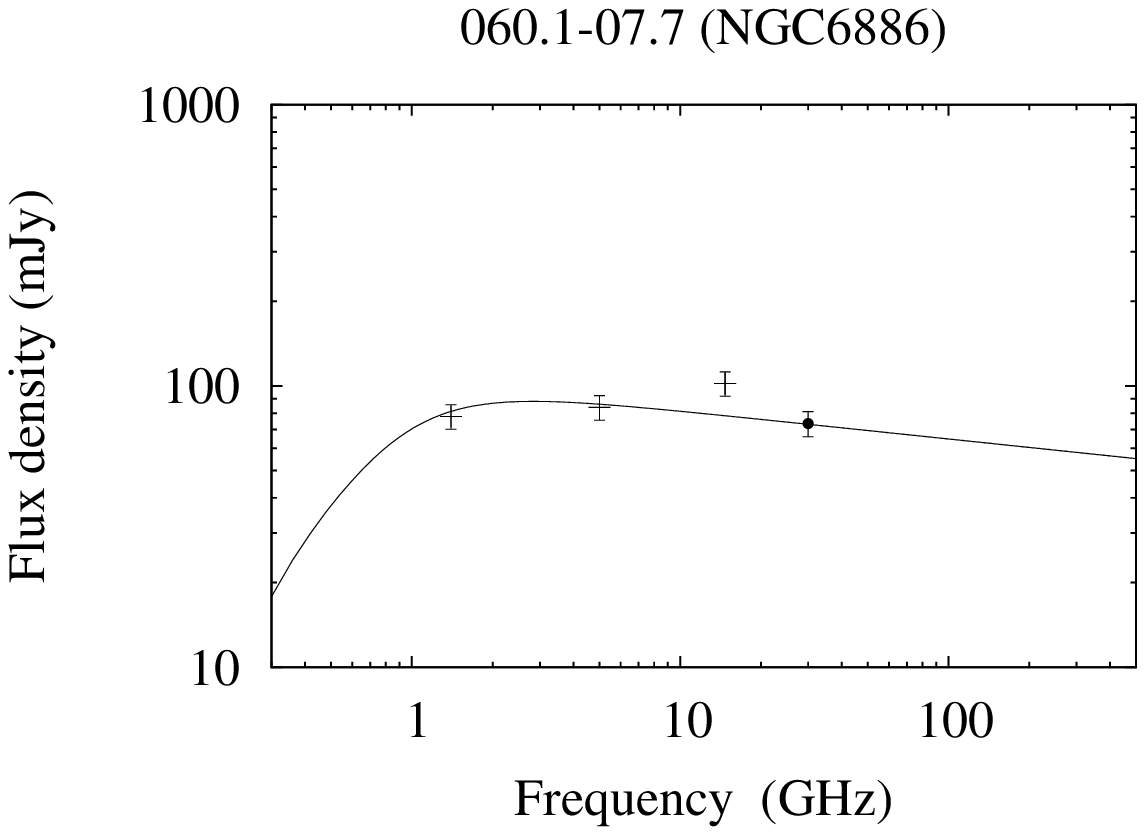}      
&
   \includegraphics[width=0.3\textwidth]{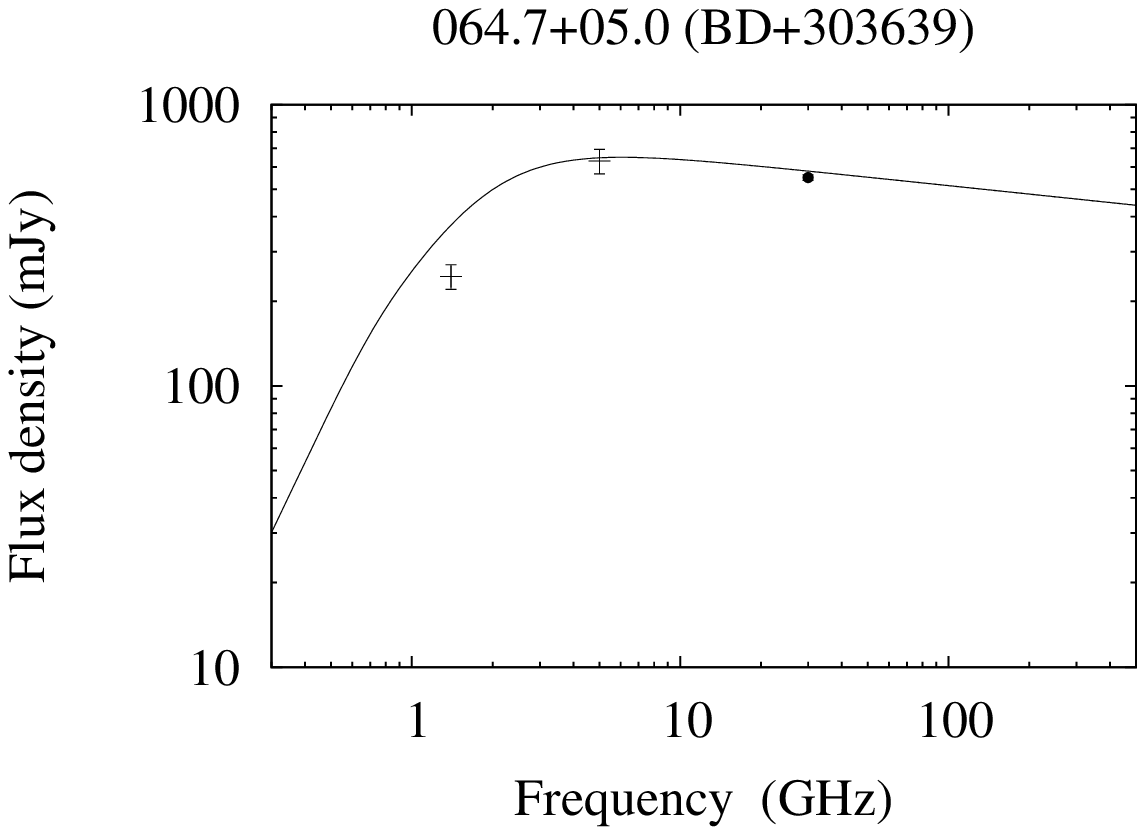}      
\\
   \includegraphics[width=0.3\textwidth]{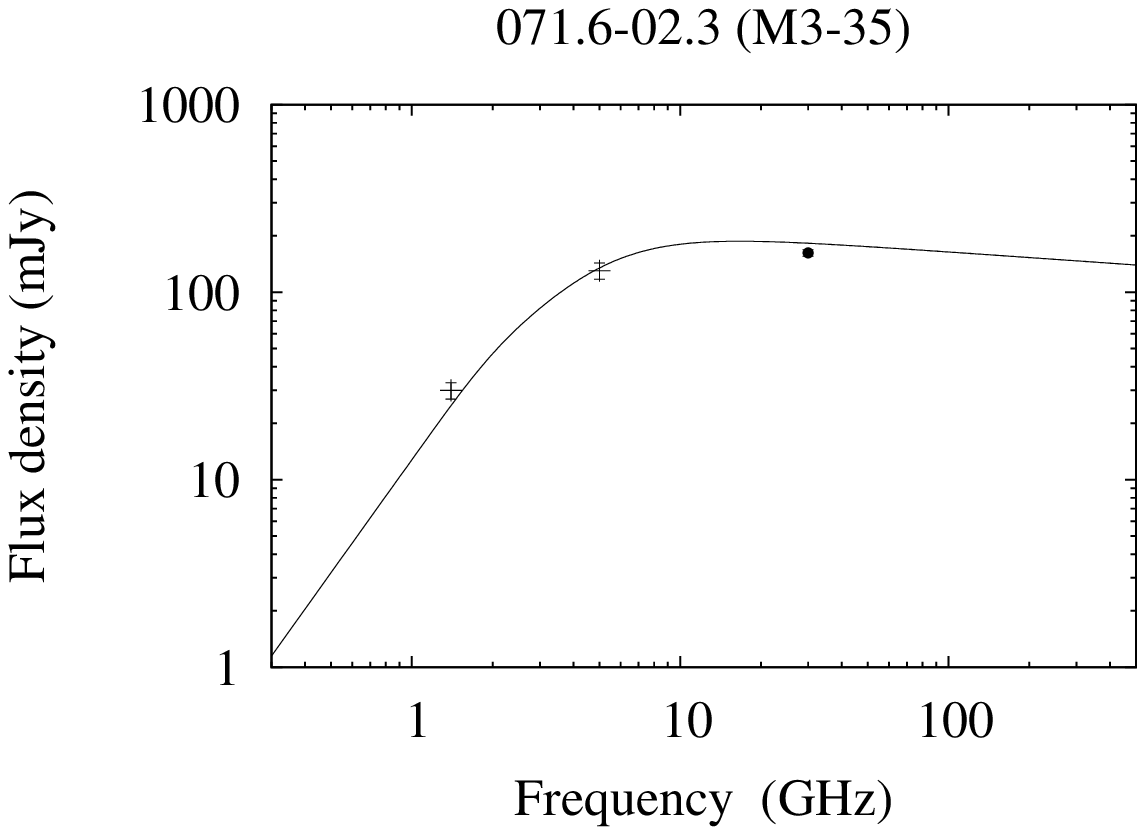}      
&
   \includegraphics[width=0.3\textwidth]{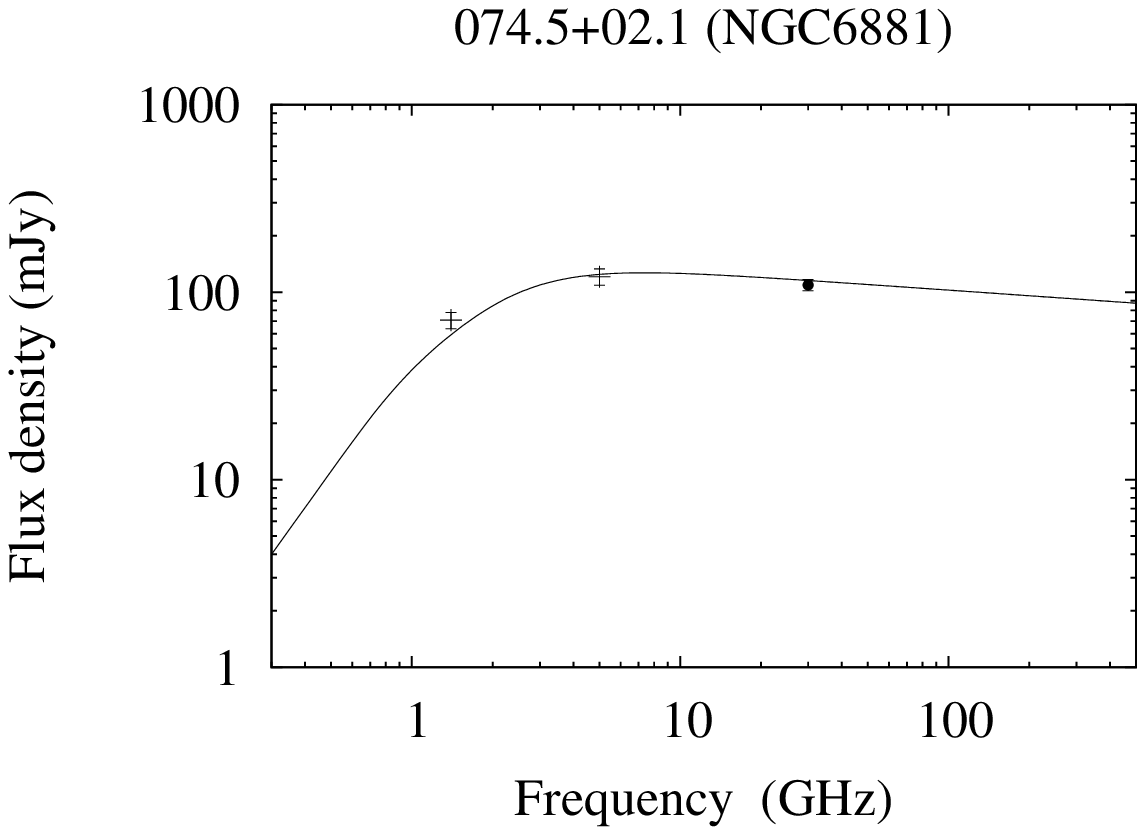}      
&
   \includegraphics[width=0.3\textwidth]{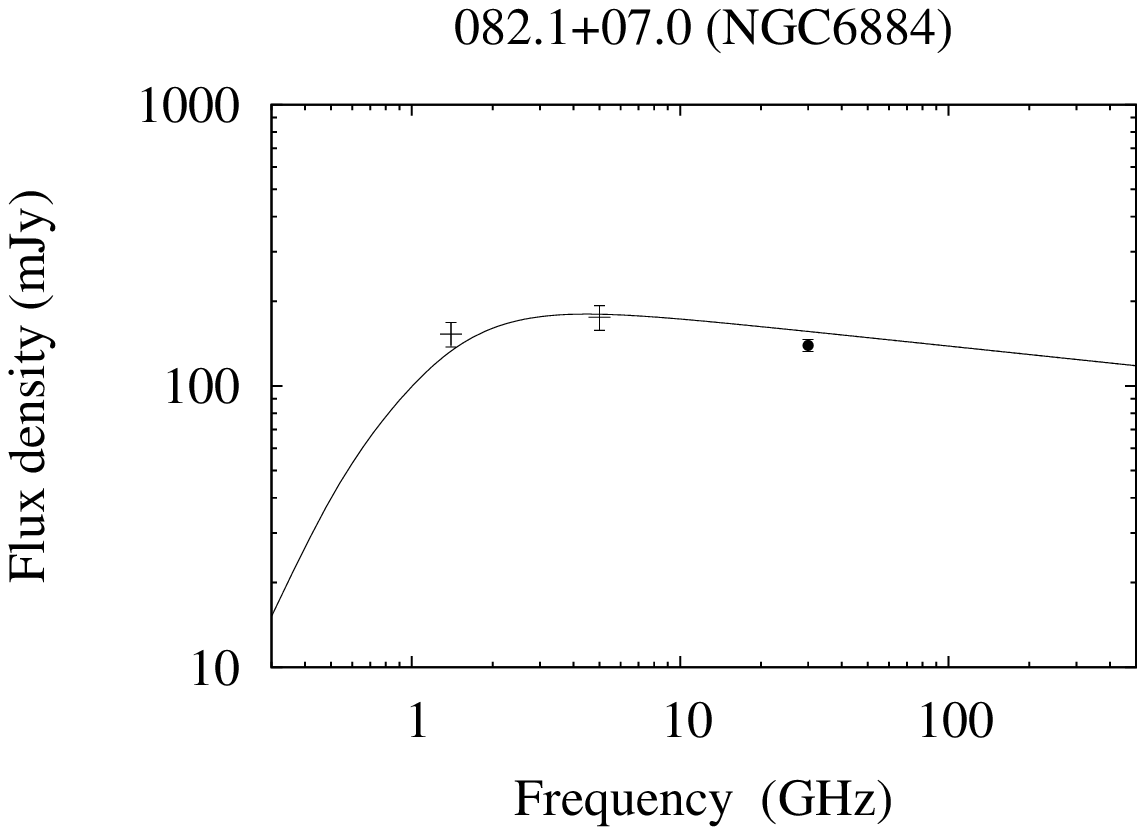}      
\\
   \includegraphics[width=0.3\textwidth]{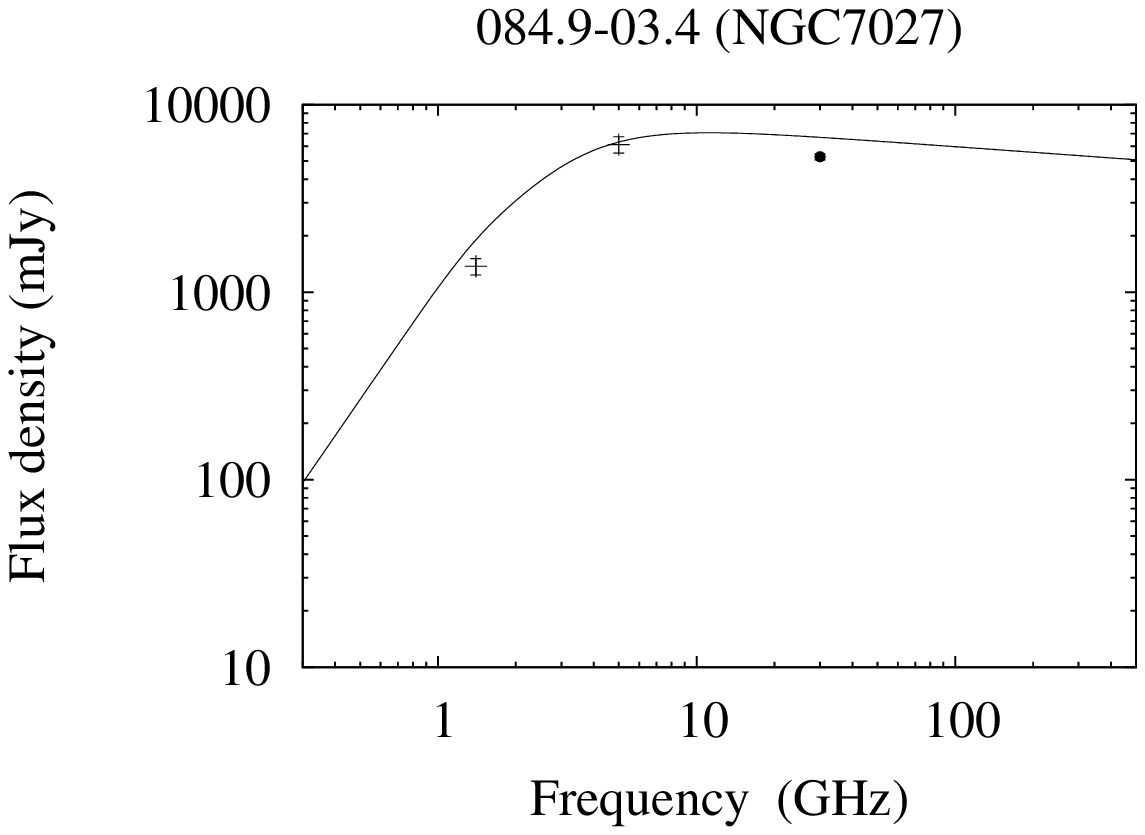}      
&
   \includegraphics[width=0.3\textwidth]{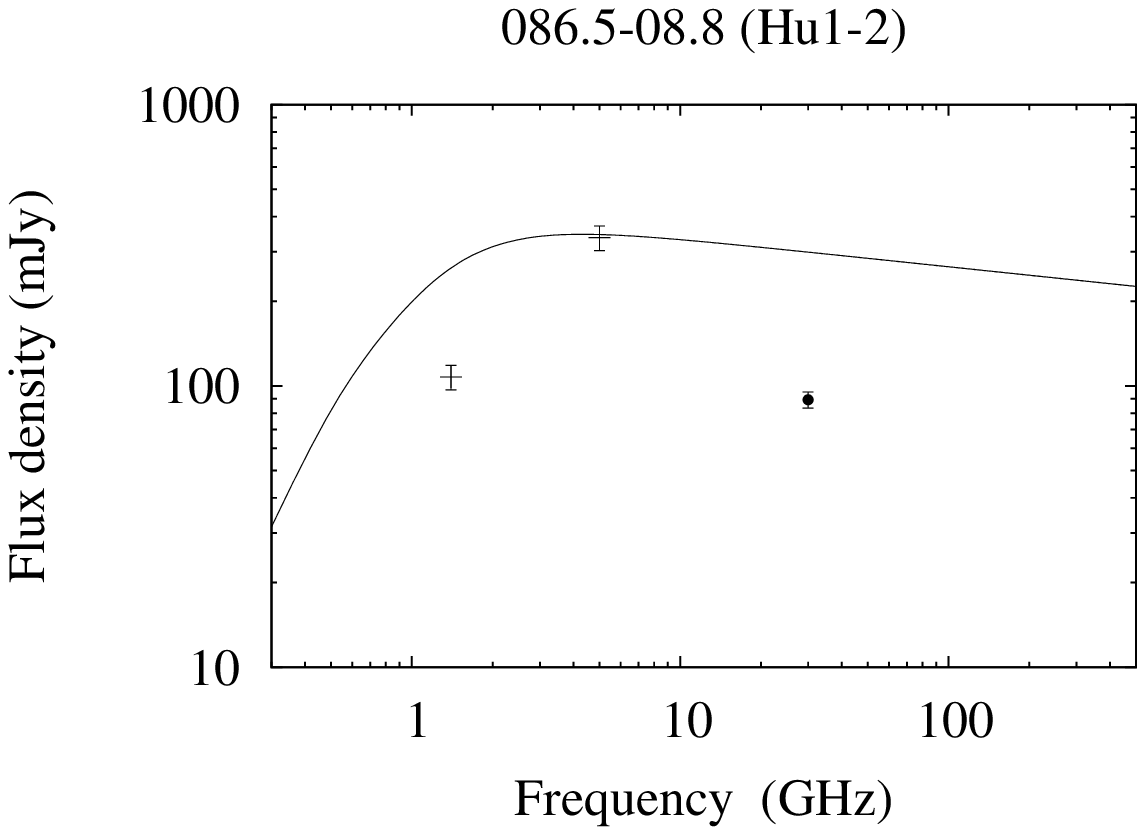}      
&
   \includegraphics[width=0.3\textwidth]{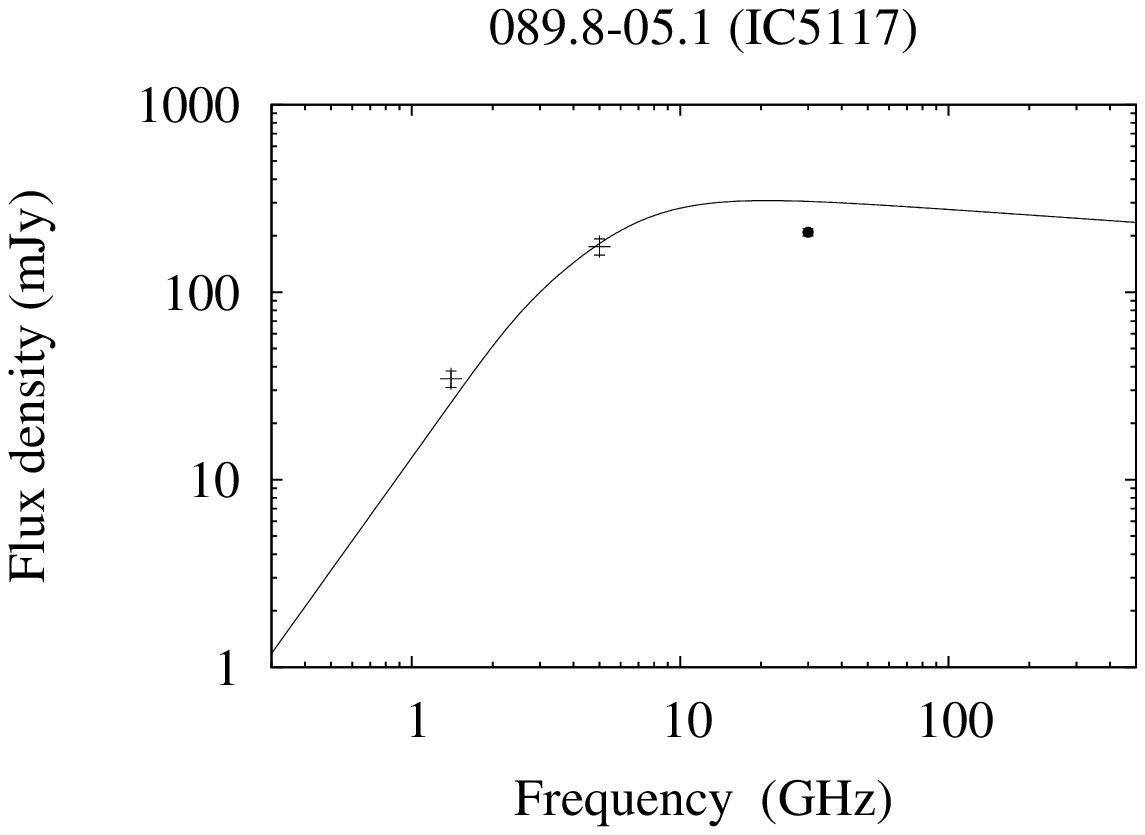}      
\\
   \includegraphics[width=0.3\textwidth]{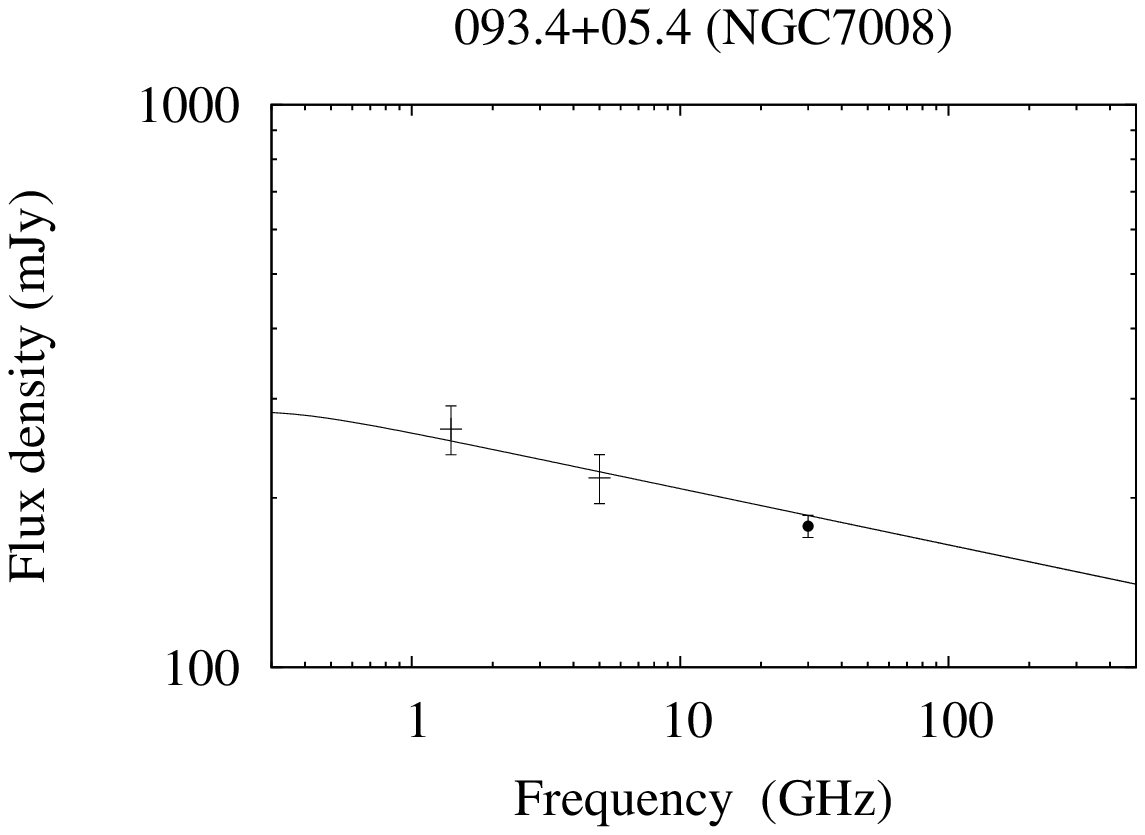}      
&
   \includegraphics[width=0.3\textwidth]{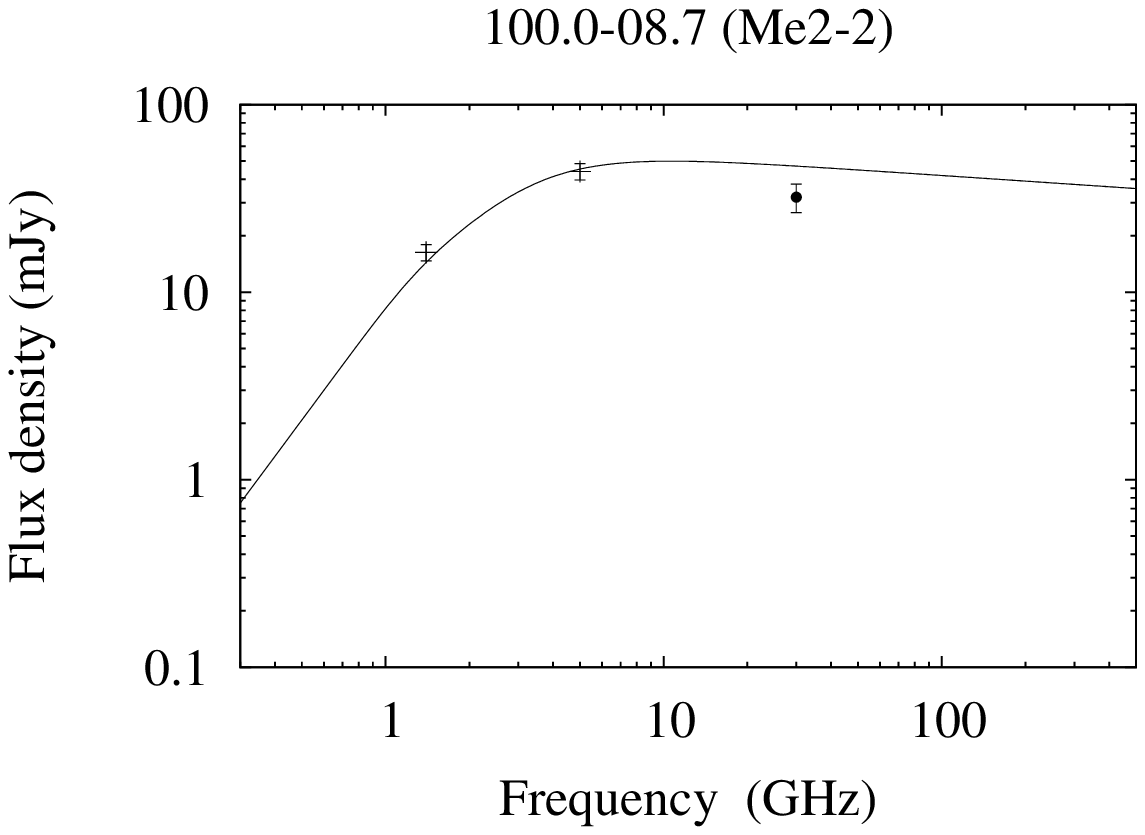}      
&
   \includegraphics[width=0.3\textwidth]{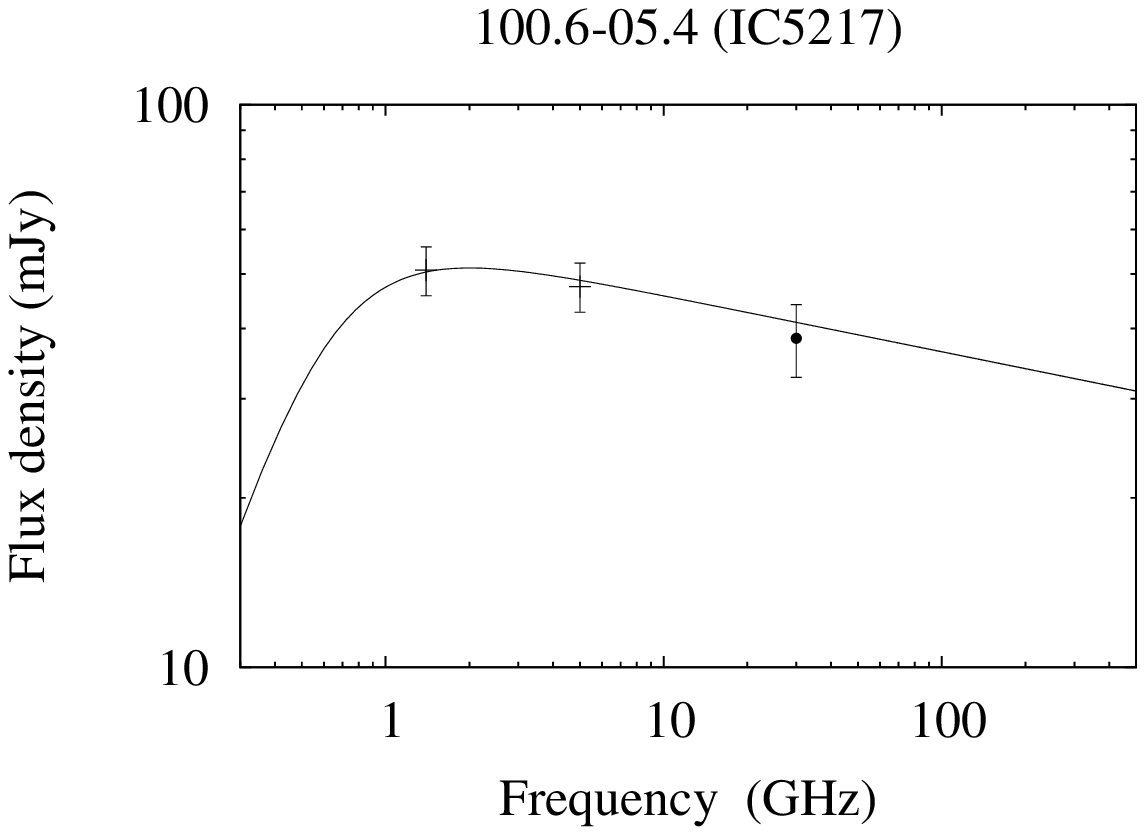}      
\\
   \includegraphics[width=0.3\textwidth]{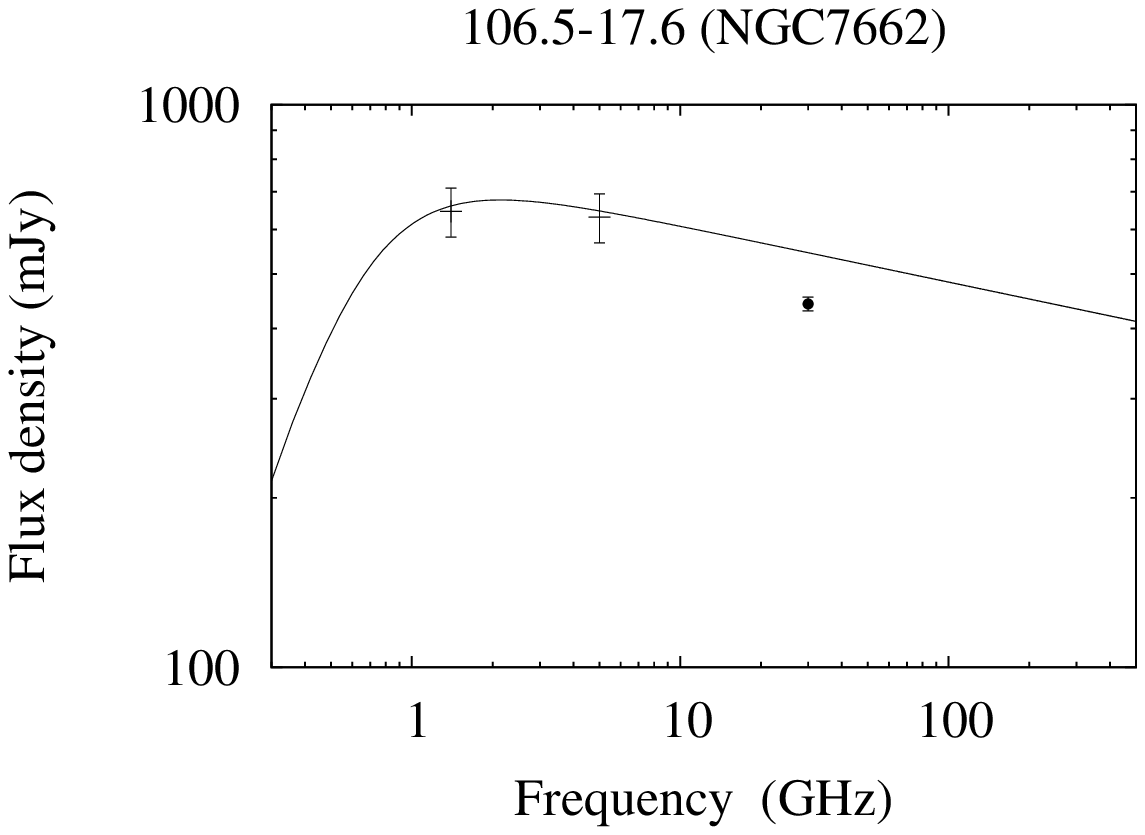}      
&
   \includegraphics[width=0.3\textwidth]{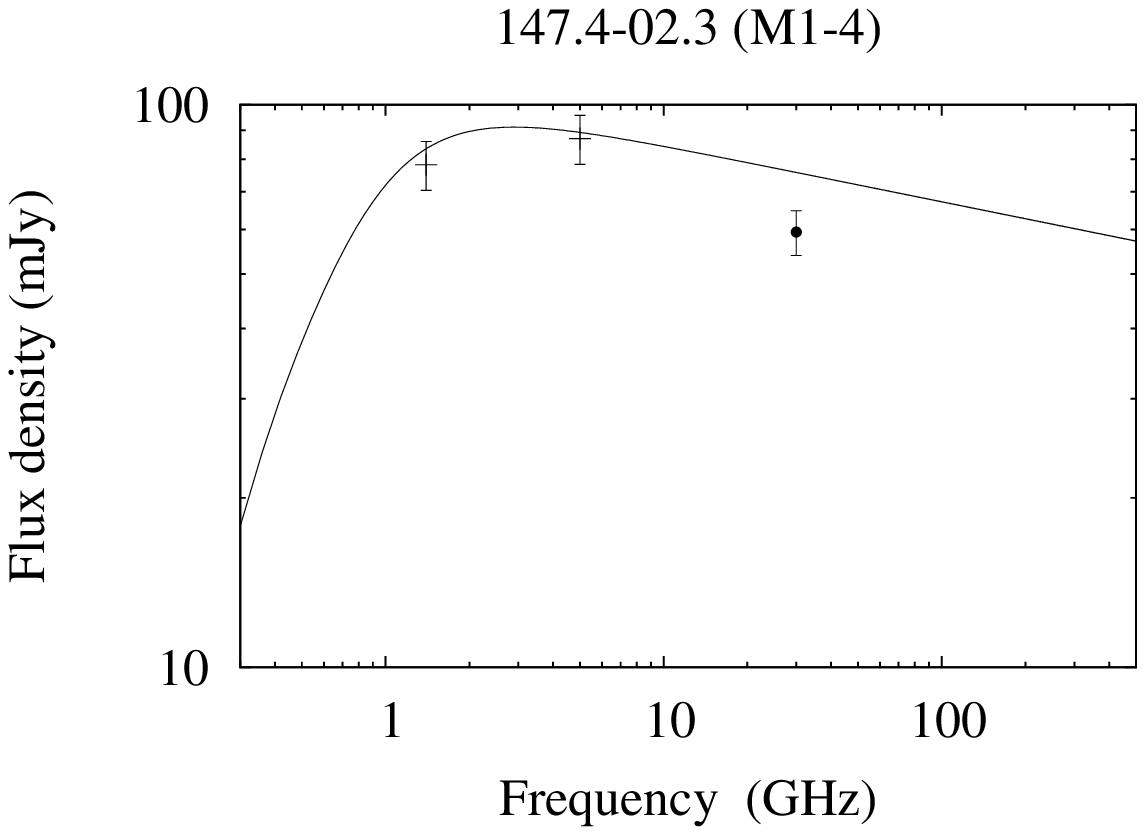}      
&
   \includegraphics[width=0.3\textwidth]{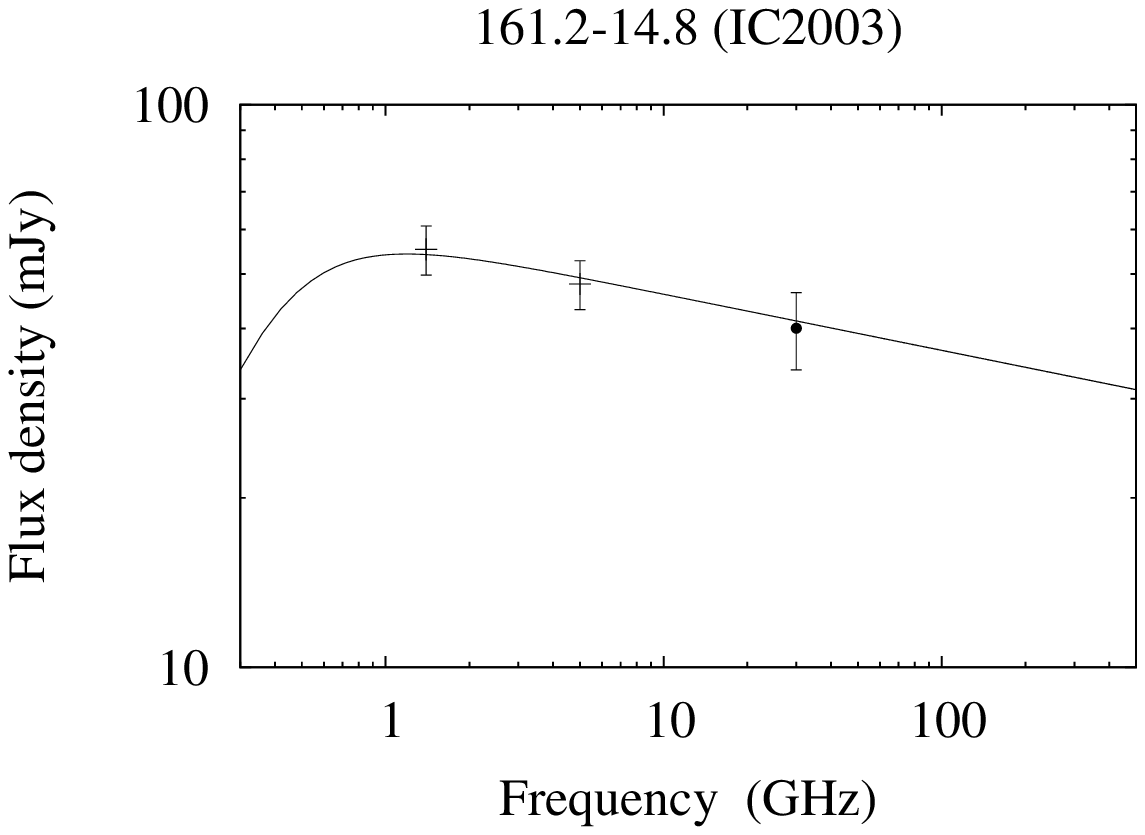}      
\\
   \includegraphics[width=0.3\textwidth]{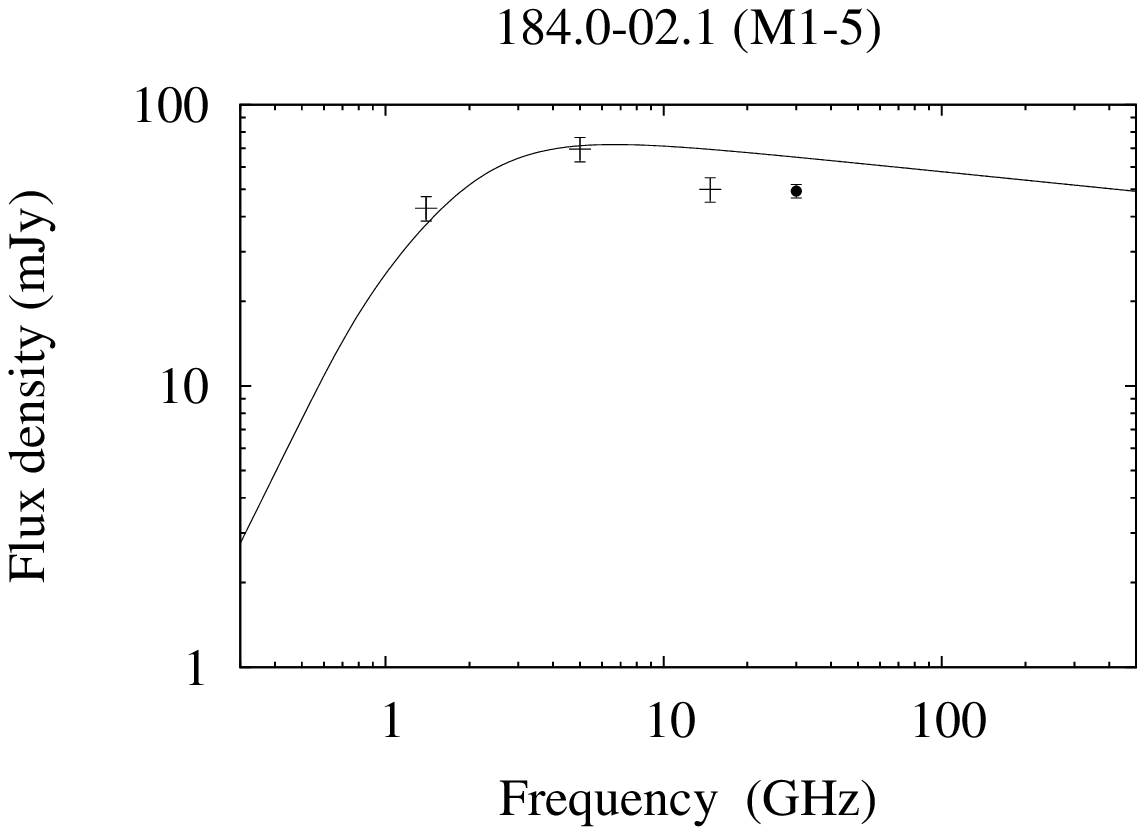}      
&
   \includegraphics[width=0.3\textwidth]{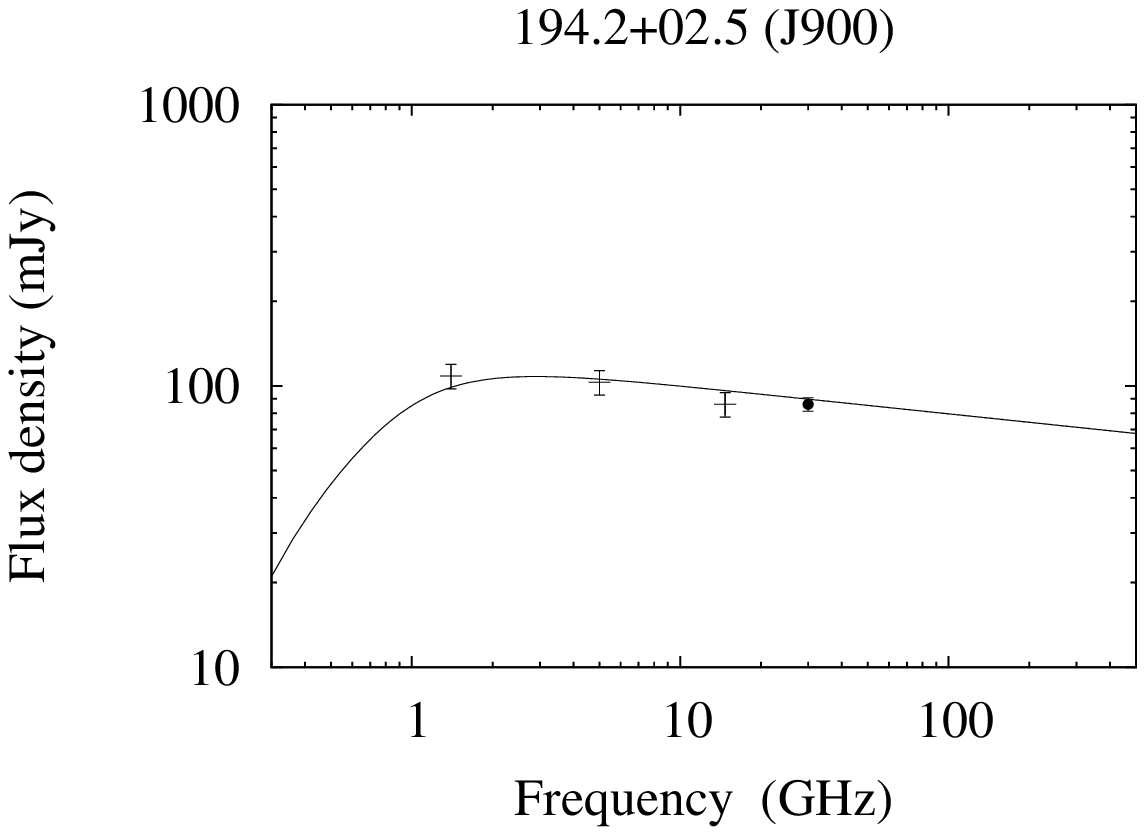}      
&
   \includegraphics[width=0.3\textwidth]{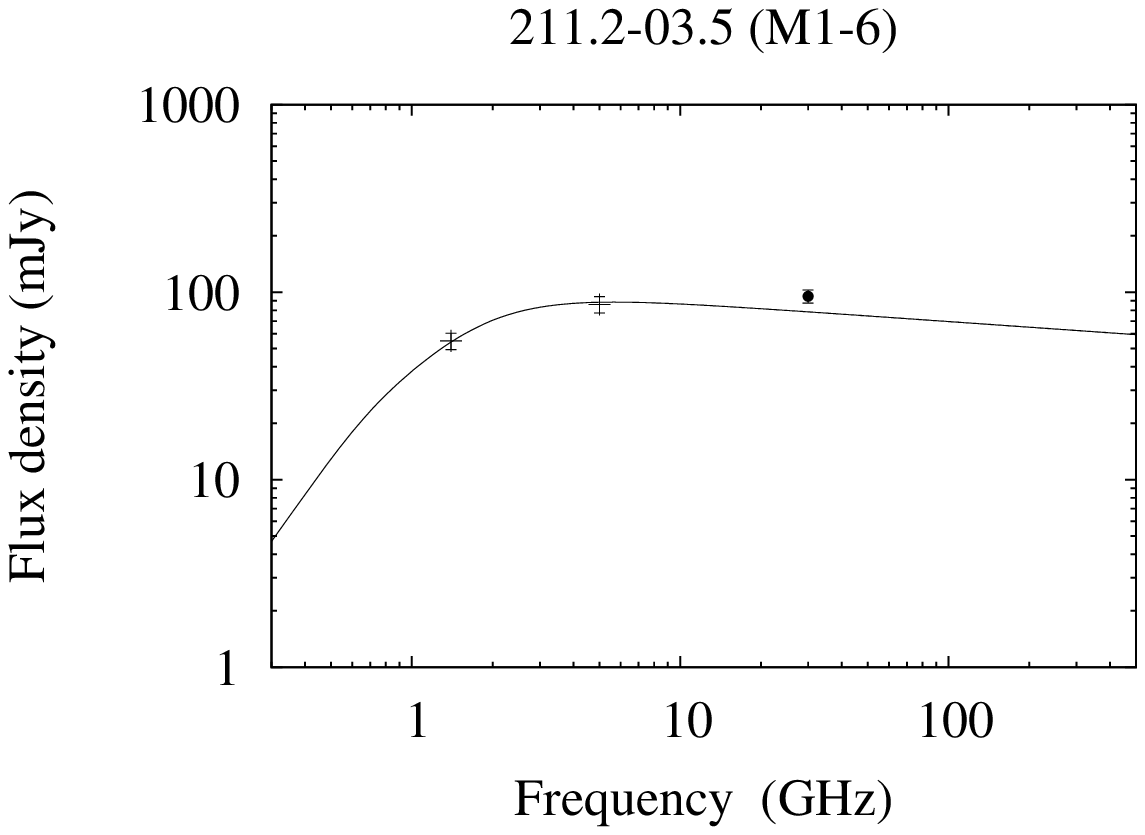}      
\\
   \includegraphics[width=0.3\textwidth]{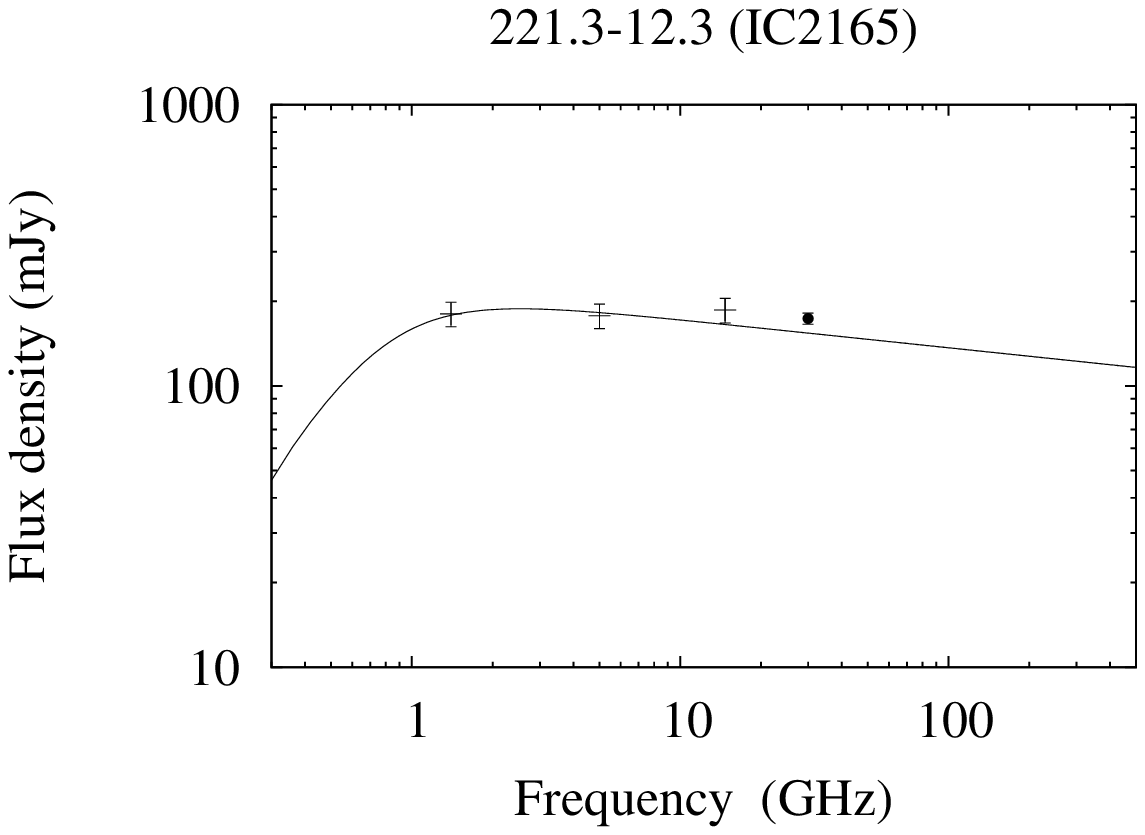}      
&
   \includegraphics[width=0.3\textwidth]{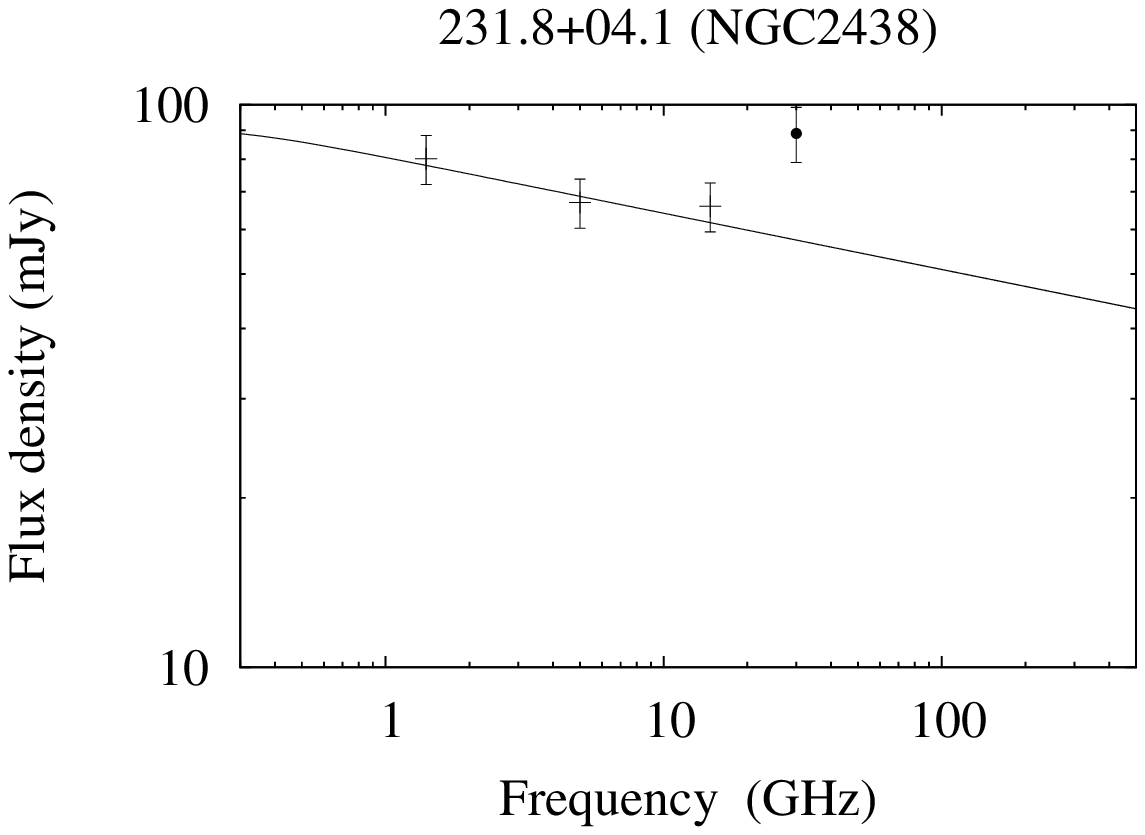}
&
\\
\end{longtable}
}

\end{document}